\newcommand{\be}{\begin{equation}}
\newcommand{\ee}{\end{equation}}
\newcommand{\brr}{\begin{eqnarray}}
\newcommand{\err}{\end{eqnarray}}
\newcommand{\nn}{\nonumber}
\newcommand{\bd}{\begin{displaymath}}
\newcommand{\ed}{\end{displaymath}}

\newcommand{\bib}{\bibitem}
\newcommand{\bfig}{\begin{figure}}
\newcommand{\efig}{\end{figure}}
\newcommand{\ie}{i.e.}
\newcommand{\eg}{e.g.}
\DeclareMathAlphabet{\mathpzc}{OT1}{pzc}{m}{it}
\newcommand{\s}[1]{{\begin{center}\colorbox{lightgray}{\begin{minipage}{.95\columnwidth}\small #1\end{minipage}}\end{center}}}
\newcommand{\h}[1]{{\begin{center}\colorbox{lime}{\begin{minipage}{.95\columnwidth}\small #1\end{minipage}}\end{center}}}
\def\alf{\alpha}
\def\bet{\beta}
\def\gam{\gamma}
\def\th{\theta}

\def\om{\omega}
\def\eps{\varepsilon}
\def\rpar{\right)}
\def\lpar{\left(}
\def\rbk{\right]}
\def\lbk{\left[}
\def\rbr{\right\}}
\def\lbr{\left\{}
\def\lb{\label}
\def\bDelta{\mbox{\boldmath $\Delta$}}
\def\ro{\mbox{\boldmath $\rho$}}

\def\non{\mbox{\boldmath $\mathcal{O}$}}
\def\fou{\mbox{\boldmath $\mathfrak{F}$}}

\def\opp{\mbox{\boldmath $\mathcal{P}$}}

\def\inc{\mbox{\scriptsize ${\rm C}$}}
\def\ins{\mbox{\scriptsize ${\rm S}$}}

\def\inn{\mbox{\tiny $\mathrm{N}$}}

\def\ina{\mbox{\tiny ${\rm A}$}}
\def\inb{\mbox{\tiny ${\rm B}$}}
\def\innu{\mbox{\tiny ${\rm U}$}}
\def\innv{\mbox{\tiny ${\rm V}$}}
\def\innx{\mbox{\tiny ${\rm X}$}}
\def\inno{\mbox{\tiny $\mathcal{O}$}}
\def\rgg{\rangle}
\def\lgg{\langle}

\def\coloneq{\mathrel{\mathop:}=}
\def\half{\frac{1}{2}}

\documentclass{elsart}
\usepackage[all]{xy}
\usepackage{ifsym}
\usepackage{pifont}
\usepackage{bbm}
\usepackage{amsfonts}
\usepackage{amsmath}
\usepackage{amsgen}
\usepackage{ifvtex}
\usepackage{ifpdf}
\usepackage[dvips]{graphicx}
\usepackage{amssymb}
\usepackage{upmath}
\usepackage{mathrsfs}
\usepackage{upgreek}
\usepackage{yfonts}
\usepackage{dsfont}
\usepackage{subfigure}
\usepackage[usenames,dvipsnames]{xcolor}
\begin{document}
\begin{frontmatter}
\title{Theoretical formulation of finite-dimensional discrete phase spaces: II. On the uncertainty principle for Schwinger unitary operators}
\author[Jabuca]{M.A. Marchiolli},
\author[Pira]{P.E.M.F. Mendon\c{c}a}
\address[Jabuca]{Avenida General Os\'{o}rio 414, centro, 14.870-100 Jaboticabal, SP, Brazil \\
                 E-mail: marcelo$\_$march@bol.com.br} 
\address[Pira]{Academia da For\c{c}a A\'{e}rea, C.P. 970, 13.643-970 Pirassununga, SP, Brazil \\
               E-mail: pmendonca@gmail.com}
\begin{abstract}
We introduce a self-consistent theoretical framework associated with the Schwinger unitary operators whose basic mathematical rules embrace a new uncertainty 
principle that generalizes and strengthens the Massar-Spindel inequality. Among other remarkable virtues, this quantum-algebraic approach exhibits a sound 
connection with the Wiener-Kinchin theorem for signal processing, which permits us to determine an effective tighter bound that not only imposes a new subtle 
set of restrictions upon the selective process of signals and wavelets bases, but also represents an important complement for property testing of unitary 
operators. Moreover, we establish a hierarchy of tighter bounds, which interpolates between the tightest bound and the Massar-Spindel inequality, as well as its 
respective link with the discrete Weyl function and tomographic reconstructions of finite quantum states. We also show how the Harper Hamiltonian and
discrete Fourier operators can be combined to construct finite ground states which yield the tightest bound of a given finite-dimensional state vector space. 
Such results touch on some fundamental questions inherent to quantum mechanics and their implications in quantum information theory.    
\end{abstract}
\end{frontmatter}
\section{Introduction}

Initially introduced by Schwinger \cite{Schwinger} for treating finite quantum systems characterized by discrete degrees of freedom immersed in a finite-dimensional
complex Hilbert space \cite{Prugo}, the unitary operators gained their first immediate application in the formal description of Pauli operators. Ever since, 
an expressive number of manuscripts \cite{Vourdas} proposed similar theoretical frameworks with intrinsic mathematical virtues and concrete applications in 
a wide family of physical systems --- here supported by a finite space of states. With regards to these state spaces, it is worth mentioning that certain algebraic
approaches related to quantum representations of finite-dimensional discrete phase spaces were constructed from this context in the past \cite{Galetti}, and tailored 
in order to properly describe the quasiprobability distribution functions \cite{Ferrie} in complete analogy with their continuous counterparts \cite{Glauber}. Thus,
applications associated with the discrete distribution functions covering different topics of particular interest in physics --- \eg, quantum information theory and
quantum computation \cite{Books,Ap1}, as well as the qualitative description of spin-tunneling effects \cite{Ap2}, open quantum systems \cite{Ap3} and magnetic 
molecules \cite{Ap4}, among others --- emerge from these approaches as a natural extension of an important robust mathematical tool. 

Although the efforts in constructing a sound theoretical framework to deal with finite-dimensional discrete phase spaces have recently achieved great advances 
(\eg, see Ref. \cite{MR1}), certain fundamental questions particularly associated with the factorization properties of finite spaces \cite{Zak}, uncertainty 
principle \cite{Massar} and property testing \cite{Wang} for the unitary operators still remain without satisfactory answers in the literature (indeed, some of
them represent open problems which do not share the same rhythm of progress). In this paper, we focus on the problem of deriving a general uncertainty principle
for Schwinger unitary operators in physics. In what follows, we discuss the relevance of such a principle and, subsequently, briefly review the results obtained 
by Massar and Spindel \cite{Massar} on this specific subject. 

To begin with, it is necessary to remember that, through an original algebraic approach which encompasses the description of finite quantum systems, Weyl \cite{Weyl} 
was the first to describe quantum kinematics as an Abelian group of ray rotations in the system space. According to Weyl: ``The kinematical structure of a
physical system is expressed by an irreducible Abelian group of unitary ray rotations in system space. The real elements of the algebra of this group are the
physical quantities of the system; the representation of the abstract group by rotations of system space associates with each such quantity a definite Hermitian
form which `represents' it." With respect to the particular case of finite state vector space, one of Weyl's most significant achievements was that the observation
of pairs of unitary rotation operators obey special commutation relations (bringing, as a result, the roots of unity) which are the unitary counterparts of the 
fundamental Heisenberg relations. Moreover, it is worth mentioning that such a ray representation of the Abelian group of rotations can be connected with some
representations of the generalized Clifford algebra, this fact being thoroughly explored by Ramakrishnan and coworkers \cite{Rama} through extensive studies of certain
physical problems. Still within the aforementioned Weyl approach for quantum kinematics, let us briefly mention that some authors have also addressed the problem of
discussing quantum mechanics in finite-dimensional state vector spaces, where the {\it coordinate} and {\it momentum} operators (characterized by discrete spectra) 
play an essential role in this context \cite{Santhanam}.

Although both Weyl and Schwinger's theoretical approaches have made seminal and complementary contributions upon the scope of unitary operators in finite physical 
systems, the relevance of a general uncertainty principle for such operators has not been clearly discussed or even mentioned with due emphasis in the past. 
Reflecting on this, Massar and Spindel \cite{Massar} have recently established a first uncertainty principle for the discrete Fourier transform \cite{Terras} whose 
range of applications in physics covers, among other topics, the Pauli operators, the {\it coordinate} and {\it momentum} operators with finite discrete spectra, the
modular variables, as well as signal processing. Furthermore, their result can also be employed to determine a modified discrete version of the
Heisenberg-Kennard-Robertson (HKR) uncertainty principle which resembles the generalized uncertainty principle (GUP) in the quantum-gravity framework \cite{MR1}. 
However, if one adopts an essentially pragmatic point of view, certain natural questions arise: ``Can Massar-Spindel inequality be recognized as a `generalized 
uncertainty principle' for all finite quantum states?" If not, ``What is the reliable starting point for obtaining a realistic description of this generalized 
uncertainty principle?"

The main goal of this paper is to present a self-consistent theoretical framework for the Schwinger unitary operators which embodies, within other virtues, an important
set of convenient inherent mathematical properties that allows us to construct suitable answers for the aforementioned questions. This theoretical framework, constituted
of numerical and analytical results, can be interpreted as a ``generalized version" of that one by Massar and Spindel, with immediate applications in quantum information 
theory and quantum computation, as well as in foundations of quantum mechanics. Next, we emphasize certain essential points of our particular construction process:
(i) Numerical computations related to a huge number $(\gtrapprox 10^{6})$ of randomly generated finite states demonstrate the existence of a nontrivial hierarchical
relation among the different bounds, the Massar-Spindel inequality being considered in such a case as a zeroth-order approximation. (ii) The existence of a tightest bound
for different dimensions of state vector space leads us to produce a sufficient number of formal results related to the Hermitian trigonometric operators (defined through
well-known specific combinations of unitary operators) and their corresponding Robertson-Schr\"{o}dinger (RS) uncertainty principles \cite{Dodonov}, which culminates in
the formulation of a new inequality which takes into account the quantum correlation effects. This tighter bound represents a new and important paradigm for signal
processing with straightforward implications on finite quantum states \cite{Wolf} and discrete approaches in GUP \cite{Berger}. (iii) Numerical and analytical approaches
\cite{Jackiw,Barker} confirm the special link between the ground state inherent to the Harper Hamiltonian and the tightest bound for any Hilbert space dimensions. 
Finally, (iv) the connection with tomographic measurements of finite quantum states via discrete Weyl function represents, in this case, a {\it tour de force} in our
investigative journey on unitary operators that allows to join both the Weyl and Schwinger quantum-algebraic approaches in an elegant way.   

This paper is structured as follows. In Section 2, we fix a preliminary mathematical background on the Schwinger unitary operators, which allows us to discuss the
implications and limitations of the Massar-Spindel inequality for signal processing. In Section 3, we introduce four Hermitian trigonometric operators through 
effective combinations of the Schwinger unitary operators. Together, these operators provide a self-consistent quantum-algebraic framework, leading us to determine 
a new tighter bound for Massar-Spindel inequality. Section 4 is dedicated to discuss certain important aspects of the tightest bounds and their respective kinematical 
link with the Harper Hamiltonian. In addition, Section 5 presents an elegant mathematical procedure for measuring a particular family of expectation values --- here 
mapped upon finite-dimensional discrete phase spaces and related to the unitary operators under investigation --- via discrete Weyl function. Section 6 contains our
summary and conclusions. Finally, Appendix A concerns the Harper functions and their respective connection with the tightest bounds verified in the numerical calculations.

\section{Preliminaries}

In order to make the presentation of this section more clear and self-contained, we begin by reviewing some essential mathematical prerequisites related 
to the Schwinger unitary operators. Only then we establish the Massar-Spindel inequality and its inherent limitations. 

\subsection{Schwinger unitary operators}

\begin{pf*}{{\bf Definition (Schwinger).}}
Let $\{ {\bf U},{\bf  V} \}$ be a pair of unitary operators defined in a $N$-dimensional state vector space, and $\{ | u_{\alf} \rgg,| v_{\bet} \rgg \}$ 
denote their respective orthonormal eigenvectors related by the inner product $\lgg u_{\alf} | v_{\bet} \rgg = \frac{1}{\sqrt{N}} \om^{\alf \bet}$ with 
$\om \coloneq \exp \lpar \frac{2 \pi \mathrm{i}}{N} \rpar$. The general properties
\bd
{\bf U}^{\eta} | u_{\alf} \rgg = \om^{\alf \eta} | u_{\alf} \rgg , \; {\bf V}^{\xi} | v_{\bet} \rgg = \om^{\bet \xi} | v_{\bet} \rgg , \;
{\bf U}^{\eta} | v_{\bet} \rgg = | v_{\bet + \eta} \rgg , \; {\bf V}^{\xi} | u_{\alf} \rgg = | u_{\alf - \xi} \rgg ,
\ed
together with the fundamental relations 
\bd
{\bf U}^{N} = {\bf 1} , \quad {\bf V}^{N} = {\bf 1} , \quad {\bf V}^{\xi} {\bf U}^{\eta} = \om^{\eta \xi} {\bf U}^{\eta} {\bf V}^{\xi} , 
\ed
constitute an important set of mathematical rules that are related to the generalized Clifford algebra \cite{Rama}. Here, the discrete labels
$\{ \alf,\bet,\eta,\xi \}$ obey the arithmetic modulo $N$ and $\lgg u_{\alf} | v_{\bet} \rgg$ represents a symmetrical finite Fourier kernel. A compilation 
of results and properties associated with ${\bf U}$ and ${\bf V}$ can be found in Ref. \cite{Schwinger}.
\end{pf*}

In the following, let $\ro$ describe a set of physical systems labeled by a finite space of states, whereas $\mathscr{V}_{\innu} \coloneq 1 - | \lgg {\bf U} \rgg |^{2}$ 
and $\mathscr{V}_{\innv} \coloneq 1 - | \lgg {\bf V} \rgg |^{2}$ denote the variances related to the respective unitary operators ${\bf U}$ and ${\bf V}$ --- in 
this case, $\lgg {\bf U} \rgg$ and $\lgg {\bf V} \rgg$ represent the mean values of ${\bf U}$ and ${\bf V}$ defined in a $N$-dimensional state vectors space. Since $\ro$
refers to a normalized density operator, the Cauchy-Schwarz inequality allows us to prove that $| \lgg {\bf U} \rgg |^{2}$ and $| \lgg {\bf V} \rgg |^{2}$ are restricted
to the closed interval $[0,1]$; consequently, both the variances are trivially bounded by $0 \leq \mathscr{V}_{\innu (\innv)} \leq 1$. Indeed, the upper and lower bounds
are promptly reached when one considers the localized bases $\lbr | u_{\alf} \rgg \lgg u_{\alf} | \rbr_{0 \leq \alf \leq N-1}$ and $\lbr | v_{\bet} \rgg \lgg v_{\bet} | 
\rbr_{0 \leq \bet \leq N-1}$, \ie, for a given $\ro = | u_{\alf} \rgg \lgg u_{\alf} | \Rightarrow \mathscr{V}_{\innu} = 0$ and $\mathscr{V}_{\innv} = 1$; otherwise, if 
$\ro = | v_{\bet} \rgg \lgg v_{\bet} | \Rightarrow \mathscr{V}_{\innu} = 1$ and $\mathscr{V}_{\innv} = 0$. Furthermore, note that $\mathscr{V}_{\innu}$ and
$\mathscr{V}_{\innv}$ are invariant under phase transformations, namely, ${\bf U} \rightarrow \mathrm{e}^{\mathrm{i} \varphi} {\bf U}$ and ${\bf V} \rightarrow
\mathrm{e}^{\mathrm{i} \theta} {\bf V}$ for any $\{ \varphi,\theta \} \in \mathbb{R}$. 

\subsection{Massar-Spindel inequality}

This inequality is based on the Wiener-Kinchin theorem for signal processing and provides a constraint between the values of $\lgg \overline{{\bf V}}^{\xi} \rgg$
(correlation function) and $\lgg \overline{{\bf U}}^{\eta} \rgg$ (discrete Fourier transform of the intensity time series). According to Massar and Spindel: 
`This kind of constraint should prove useful in signal processing, as it constrains what kinds of signals are possible, or what kind of wavelet bases one can construct.'
We state this result in the theorem below (proved in the supplementary material from Ref. \cite{Massar}), for then proceeding with a numerical study of its content
and first implications. 

\s{\begin{pf*}{{\bf Theorem (Massar and Spindel).}}
{\it Let $\overline{{\bf U}}$ and $\overline{{\bf V}}$ denote two unitary operators such that $\overline{{\bf U}} \, \overline{{\bf V}} = \mathrm{e}^{\mathrm{i} \Phi} 
\overline{{\bf V}} \, \overline{{\bf U}}$ and $\overline{{\bf U}}^{\dagger} \overline{{\bf V}} = \mathrm{e}^{- \mathrm{i} \Phi} \overline{{\bf V}} \, 
\overline{{\bf U}}^{\dagger}$ with $\Phi \in [0,\pi)$. The variances $\mathscr{V}_{\overline{\innu}}$ and $\mathscr{V}_{\overline{\innv}}$ --- here defined for a given
quantum state $\ro$ and limited to the closed interval $[0,1]$ --- satisfy the inequality
\be
\lb{s2e1}
(1+2A) \mathscr{V}_{\overline{\innu}} \mathscr{V}_{\overline{\innv}} \geq A^{2} (1 - \mathscr{V}_{\overline{\innu}} - \mathscr{V}_{\overline{\innv}})
\ee
where $A = \tan \lpar \frac{\Phi}{2} \rpar$. The saturation is reached for localized bases.}
\end{pf*}}

This theorem leads us, in principle, to consider the different possibilities of connections between the Schwinger unitary operators $\{ {\bf U},{\bf V} \}$ 
(defined in the previous subsection) and $\{ \overline{\bf U},\overline{\bf V} \}$. A first immediate link yields the relation $\overline{\bf U} = {\bf U}$ and 
$\overline{\bf V} = {\bf V}$ for $\Phi = - \frac{2 \pi}{N}$, which implies in the apparent violation of Eq. ({\ref{s2e1}) since $\Phi \notin [0,\pi)$. The second 
connection establishes the alternative relation $\overline{\bf U} = {\bf V}$ and $\overline{\bf V} = {\bf U}$ with $\Phi = \frac{2 \pi}{N}$, this result being 
responsible for validating the Massar-Spindel inequality. It is important to mention that the apparent problem detected in the first situation can be properly 
circumvented making $\Phi \rightarrow - \Phi$ and $A \rightarrow -A$ with $\Phi = \frac{2 \pi}{N}$ fixed. That is, the modified bound 
\bd
(1-2A) \mathscr{V}_{\innu} \mathscr{V}_{\innv} \geq A^{2} (1 - \mathscr{V}_{\innu} - \mathscr{V}_{\innv})
\ed
holds for any $A = \tan \lpar \frac{\pi}{N} \rpar$ and integer $N \in [2,\infty)$.  

For the sake of simplicity and convenience, let us now introduce the shift operator $\Delta \non \equiv \non - \lgg \non \rgg$ with $\lgg \non \rgg \neq 0$ for 
a given arbitrary unitary operator $\non$ and density operator $\ro$, which leads to define\footnote{It is important to note a certain level of arbitrariness in 
this definition because there is no especification whatsoever of which density operator $\ro$ is used to evaluate the mean value in the denominator of Eq. (\ref{s2e2}).
Henceforth, this arbitrariness will be removed by exploring the fact that, in our computations, the operator $\delta \non$ will always appear as an argument of a 
variance function --- in such a situation, we will understand that the expectation value $\lgg \non \rgg$ is computed with respect to the same state used in the
computation of $\lgg \delta \non \rgg$.} its non-unitary counterpart as follows \cite{Leo}:
\be
\lb{s2e2}
\delta \non \coloneq \frac{\Delta \non}{\lgg \non \rgg} = \frac{\non - \lgg \non \rgg}{\lgg \non \rgg} .
\ee
Following, it is straighforward to show that both the variances $\mathscr{V}_{\inno}$ and $\mathscr{V}_{\delta \inno}$ are related through the expressions 
\be
\lb{s2e3}
\mathscr{V}_{\delta \inno} = \frac{\mathscr{V}_{\inno}}{1 - \mathscr{V}_{\inno}} \; \; \lpar 0 \leq \mathscr{V}_{\delta \inno} < \infty \rpar \qquad 
\mbox{and} \qquad \mathscr{V}_{\inno} = \frac{\mathscr{V}_{\delta \inno}}{1 + \mathscr{V}_{\delta \inno}} ,
\ee
which substituted into inequality (\ref{s2e1}) for $\overline{\bf U} = {\bf V}$ and $\overline{\bf V} = {\bf U}$ gives
\be
\lb{s2e4}
\mathscr{V}_{\delta \innu} \mathscr{V}_{\delta \innv} \geq \eps^{2} \qquad \mbox{with} \qquad \eps = \frac{A}{1+A} .
\ee
\begin{figure}[!th]
\centering
\begin{minipage}[b]{0.40\linewidth}
\includegraphics[width=\textwidth]{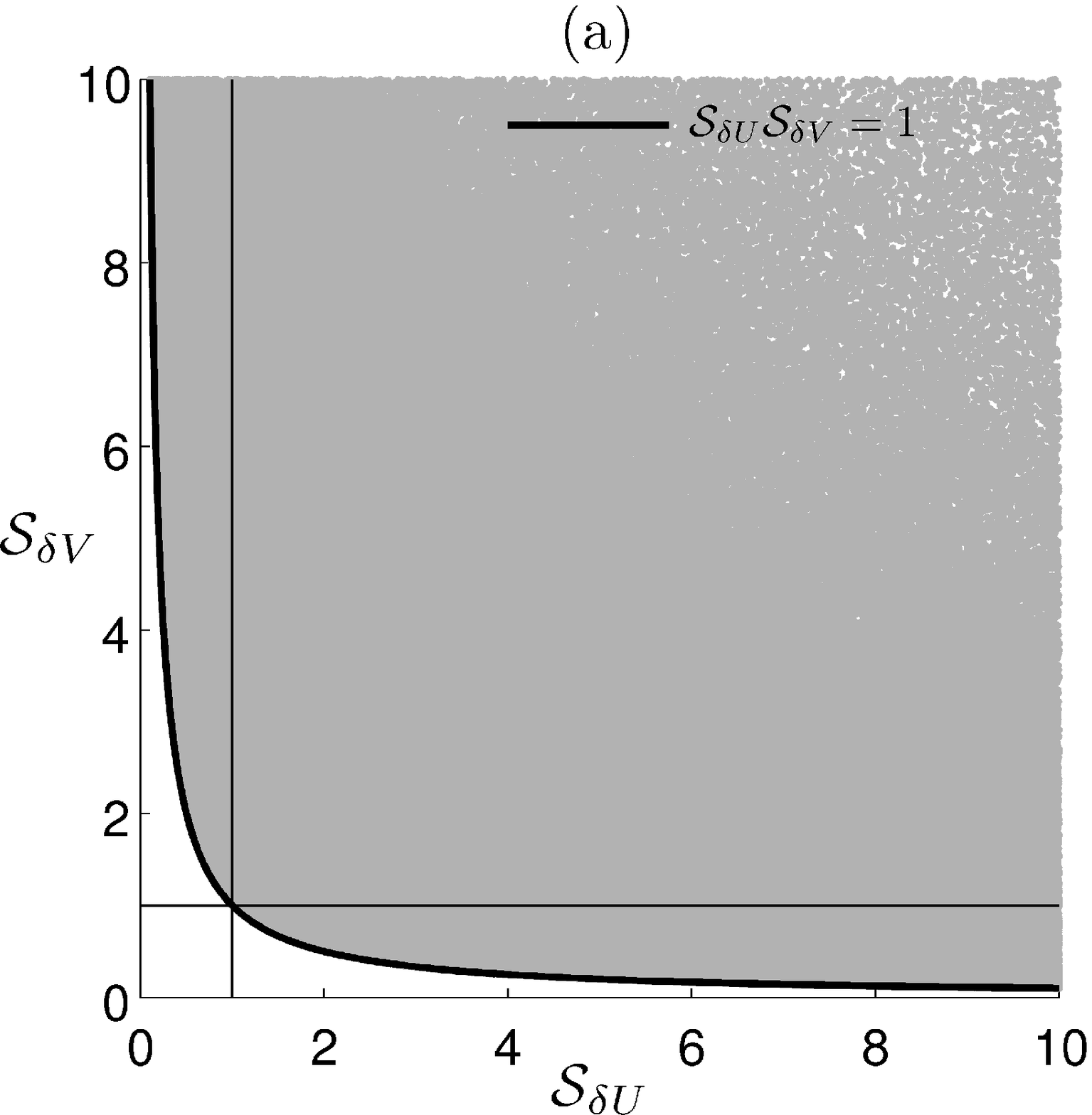}
\end{minipage} \hfill
\begin{minipage}[b]{0.40\linewidth}
\includegraphics[width=\textwidth]{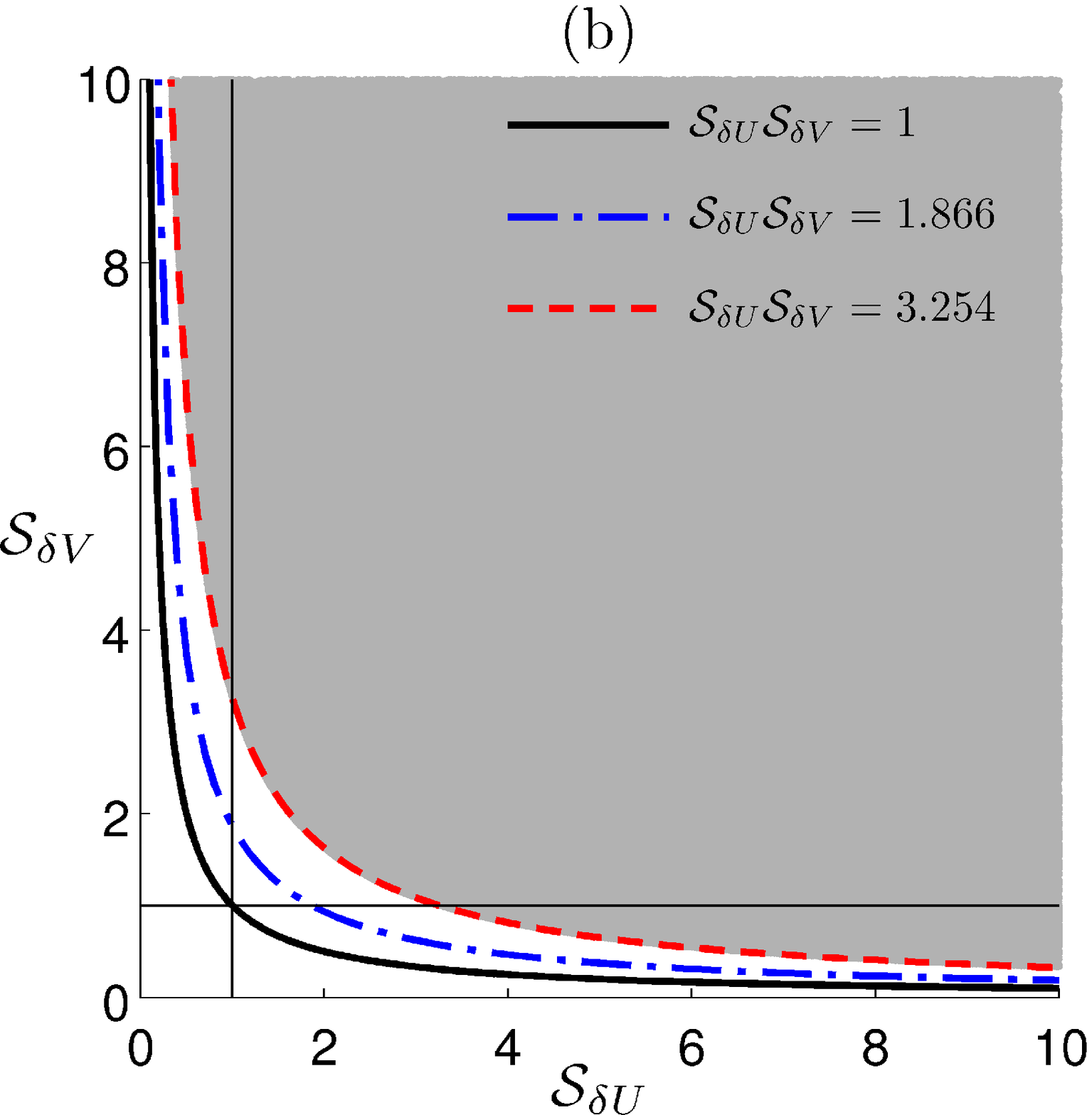}
\end{minipage} \hfill \\ \vspace*{2mm}
\begin{minipage}[b]{0.40\linewidth}
\includegraphics[width=\textwidth]{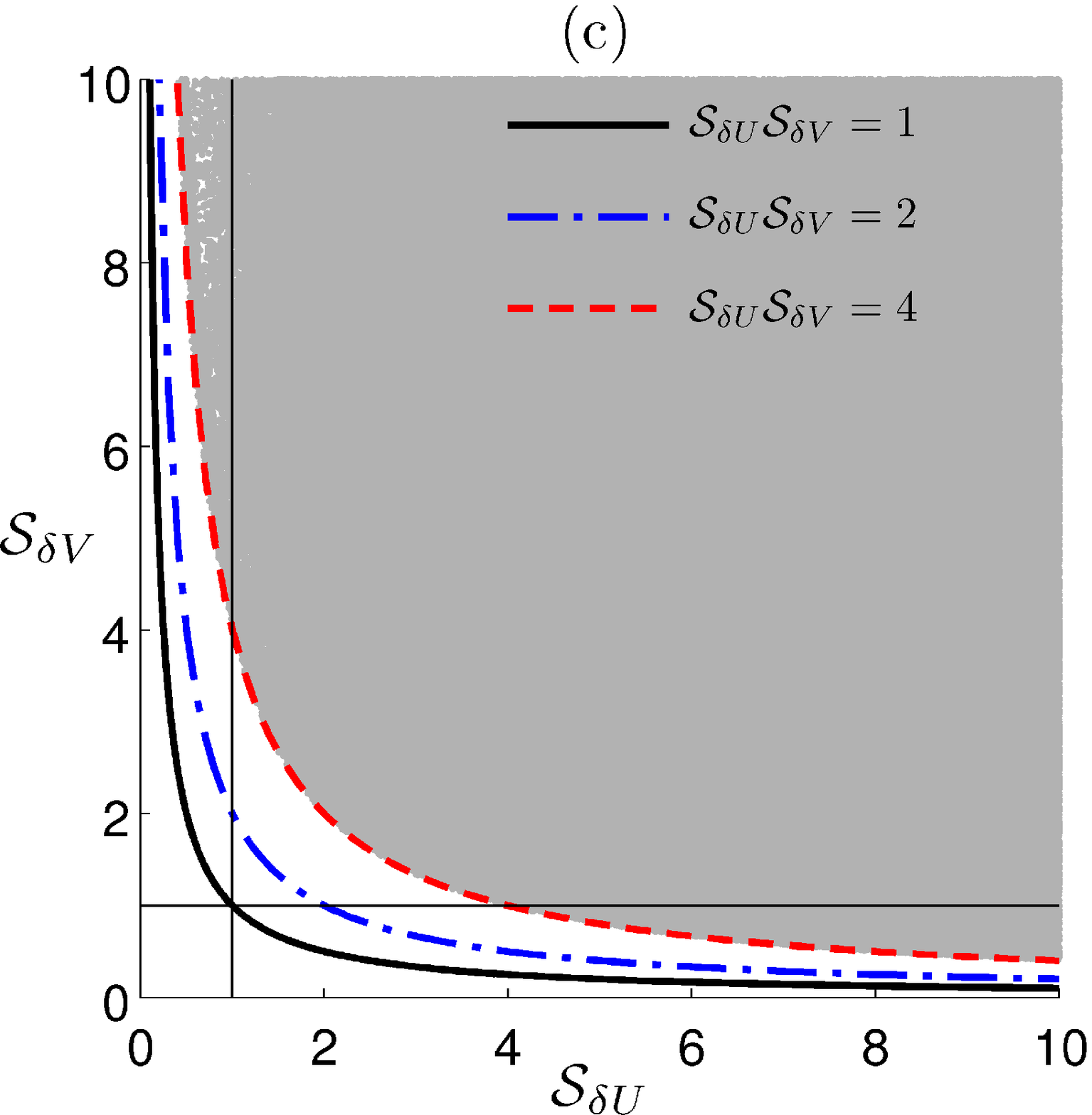}
\end{minipage} \hfill
\begin{minipage}[b]{0.40\linewidth}
\includegraphics[width=\textwidth]{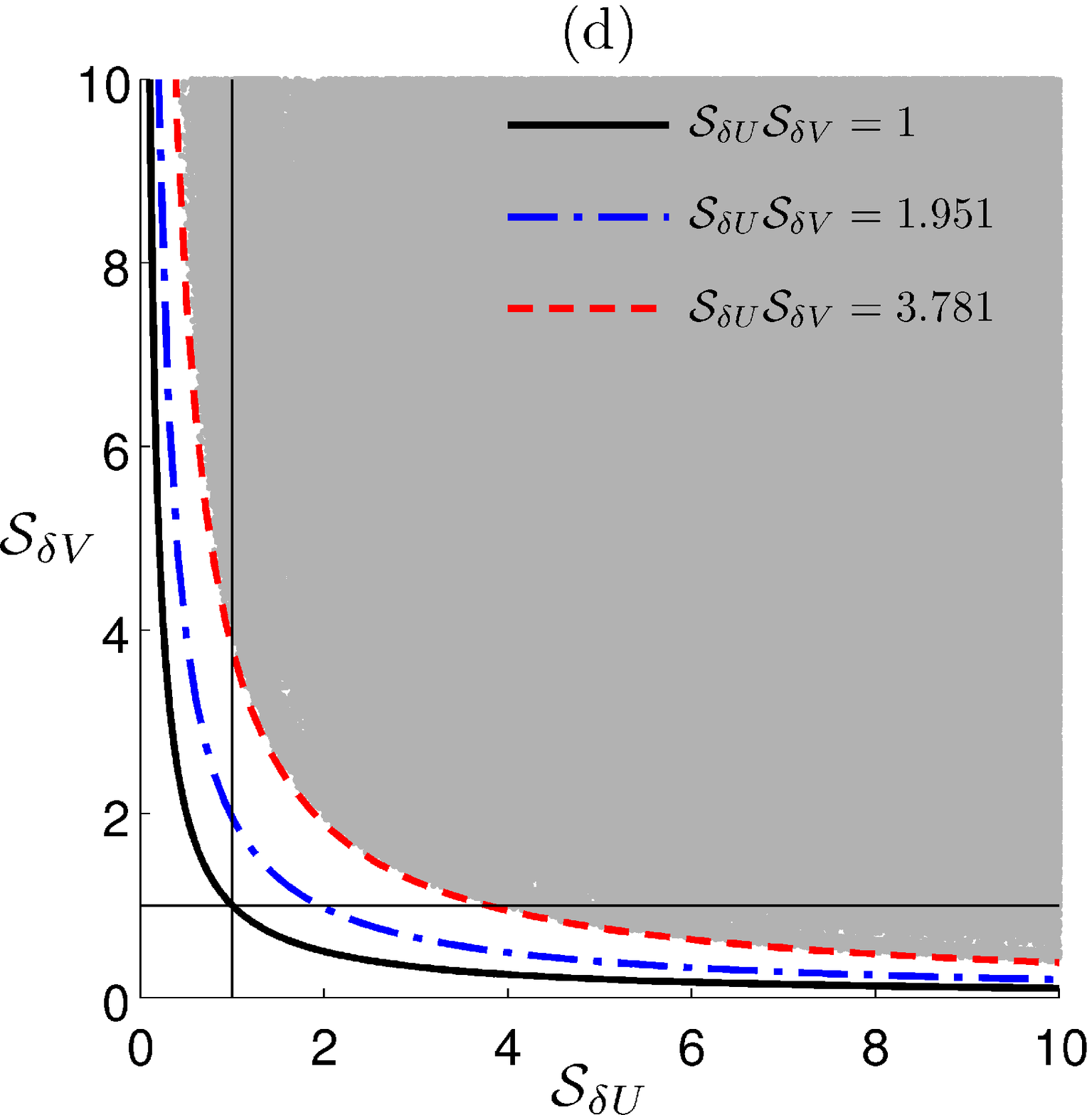}
\end{minipage} 
\caption{Plots of $\mathcal{S}_{\delta \innv}$ versus $\mathcal{S}_{\delta \innu}$ for different values of $N$ --- $2$ (a), $3$ (b), $4$ (c), and $5$ (d) --- with
approximately $10^{6}$ randomly generated states, illustrating the Massar-Spindel inequality $\lpar \mathcal{S}_{\delta \innu} \mathcal{S}_{\delta \innv} \geq 1 \rpar$
for the Schwinger unitary operators. Note that $0< \mathcal{S}_{\delta \innu}, \mathcal{S}_{\delta \innv} \leq 10$ characterizes a particular visualization window which
leads us to guess the existence of a tightest bound (see dashed curves) for each value of dimension $N$, since the distance between the saturation curve
$\mathcal{S}_{\delta \innu} \mathcal{S}_{\delta \innv} = 1$ (solid line) and the cloud of points --- generated by numerical calculations --- increases for $N > 2$. 
The dot-dashed curves showed in pictures (b,c,d) represent the intermediate inequality $\mathcal{S}_{\delta \innu} \mathcal{S}_{\delta \innv} \geq 1 + \sin \lpar 
\frac{2 \pi}{N} \rpar$, this result being considered as a first approximation to our initial intents.} 
\end{figure}

To illustrate Eq. (\ref{s2e4}) and corroborate the analytic results obtained by Massar and Spindel, let us introduce the parameters $\mathcal{S}_{\delta \innu} =
\eps^{-1} \mathscr{V}_{\delta \innu}$ and $\mathcal{S}_{\delta \innv} = \eps^{-1} \mathscr{V}_{\delta \innv}$ such that $\mathcal{S}_{\delta \innu} 
\mathcal{S}_{\delta \innv} \geq 1$. This particular inequality defines a region in the two-dimensional space limited by the rectangular hyperbola 
$\mathcal{S}_{\delta \innv} = \mathcal{S}_{\delta \innu}^{-1}$ that preserves the original equation $\mathscr{V}_{\delta \innu} \mathscr{V}_{\delta \innv} \geq \eps^{2}$.
Figure 1 shows the plots of $\mathcal{S}_{\delta \innv}$ versus $\mathcal{S}_{\delta \innu}$ for approximately $10^{6}$ states of randomly generated $\ro$, within the
visualization window $0 < \mathcal{S}_{\delta \innu} , \mathcal{S}_{\delta \innv} \leq 10$, with (a) $N=2$ $(\eps \rightarrow 1)$, (b) $N=3$, (c) $N=4$, and (d) $N=5$.
Note that, excepting picture (a), all the subsequent cases exhibit a gap between the distribution of states and the rectangular hyperbola, the size of such a gap
being dependent on the value of $N$ (such an evidence motivates the search for a tightest bound that corroborates the numerical calculations). In fact, the dashed lines
in (b,c,d) describe hyperbolic curves of the form $\mathcal{S}_{\delta \innu} \mathcal{S}_{\delta \innv} = x$, where the value of $x$ was chosen as the smallest value
of $\mathcal{S}_{\delta \innu} \mathcal{S}_{\delta \innv}$ amongst the ones computed with the randomly sampled states. Later in this paper, a more rigorous procedure
for obtaining such values will be outlined. Finally, let us briefly mention that the dot-dashed lines exhibited in these pictures correspond to the intermediate result
$\mathcal{S}_{\delta \innu} \mathcal{S}_{\delta \innv} \geq 1 + \sin \lpar \frac{2 \pi}{N} \rpar$, which will be properly demonstrated in the next section.

\section{A Hierarchy of Tighter Bounds}

In the first part of this paper, we established a basic theoretical framework related to the Schwinger unitary operators where the Massar-Spindel inequality 
and its inherent limitations occupied an important place in our discussion on uncertainty principles for physical systems labeled by a finite space of states. 
At this moment, let us clarify some fundamental points raised by those results: (i) the aforementioned state spaces consist of $N$-dimensional Hilbert spaces; 
(ii) the Massar-Spindel inequality represents a ``zeroth-order approximation" in the hierarchy of uncertainty principles; and finally, (iii) the results obtained
from the numerical calculations reveal certain unexplored intrinsic properties of some finite quantum states \cite{MRG-2007,Ingemar} with potential applications 
in quantum information theory and quantum computation. In this second part, we begin the construction of a solid algebraic framework based on the RS uncertainty 
principle, which leads us, in a first moment, to determine a self-consistent set of results for the unitary operators ${\bf U}$ and ${\bf V}$ that permits to 
generalize the Massar-Spindel inequality. Indeed, these results represent an important tool in our search for tighter bounds (see numerical evidence exhibited 
in Fig. 1), whose intermediate uncertainty principles will constitute a hierarchical relation between the Massar-Spindel inequality and the tightest bound.

\subsection{Quantum-algebraic framework}

Let $\{ {\bf A},{\bf B} \}$ denote a pair of Hermitian operators defined in a $N$-dimensional state vectors space which obey the RS uncertainty principle \cite{Dodonov}
\be
\lb{s3e5}
\mathscr{V}_{\ina} \mathscr{V}_{\inb} \geq \mathscr{C}^{2}({\bf A},{\bf B}) + \frac{1}{4} \left| \lgg [ {\bf A},{\bf B} ] \rgg \right|^{2} ,
\ee
where $\mathscr{C}({\bf A},{\bf B}) \coloneq \lgg \frac{1}{2} \{ {\bf A},{\bf B} \} \rgg - \lgg {\bf A} \rgg \lgg {\bf B} \rgg$ represents the covariance function,
and $[ {\bf A},{\bf B} ]$ $\lpar \{ {\bf A},{\bf B} \} \rpar$ corresponds to the commutator (anticommutator) between ${\bf A}$ and ${\bf B}$.\footnote{It is 
important to emphasize that, according to Cauchy-Schwarz inequality, the covariance function $\mathscr{C}({\bf A},{\bf B})$ is restricted to the closed interval 
$\lbk - \sqrt{\mathscr{V}_{\ina} \mathscr{V}_{\inb}} , \sqrt{\mathscr{V}_{\ina} \mathscr{V}_{\inb}} \, \rbk$, namely, $\left| \mathscr{C}({\bf A},{\bf B}) \right| 
\leq \sqrt{\mathscr{V}_{\ina} \mathscr{V}_{\inb}}$. Indeed, for a given operator ${\bf C} \coloneq {\bf A} - \frac{\mathscr{C}({\bf A},{\bf B})}{2 \mathscr{V}_{\inb}} 
{\bf B}$ with $\mathscr{V}_{\inb} \neq 0$ and $\mathscr{V}_{\inc} \geq 0$, it turns immediate to obtain the relation $\mathscr{V}_{\inc} = \mathscr{V}_{\ina} -
\frac{\mathscr{C}^{2}({\bf A},{\bf B})}{\mathscr{V}_{\inb}}$, which demonstrates the previous result for $\mathscr{C}({\bf A},{\bf B})$. Moreover, let us also state 
a very useful result for both the commutation and anticommutation relations between ${\bf A}$ and ${\bf B}$, that is $| \lgg [ {\bf A},{\bf B} ] \rgg |^{2} + | \lgg 
\{ {\bf A},{\bf B} \} \rgg |^{2} = 4 | \lgg {\bf A} {\bf B} \rgg |^{2}$.} Next, since ${\bf U}$ and ${\bf V}$ are generally non-Hermitian operators, let us consider 
the cartesian decomposition of an arbitrary non-Hermitian operator $\non$ into its `real' and `imaginary' parts as follows \cite{Bhatia}:
\bd
\non = \lpar \frac{\non + \non^{\dagger}}{2} \rpar + \mathrm{i} \lpar \frac{\non - \non^{\dagger}}{2 \mathrm{i}} \rpar = \mathrm{Re} [\non] + \mathrm{i} \, \mathrm{Im} [\non] .
\ed
Note that $\mathrm{Re} [\non]$ and $\mathrm{Im} [\non]$ represent two Hermitian operators that comply with the RS uncertainty principle and allow to introduce, in
particular, the cosine and sine operators through the Schwinger unitary operators. 

\begin{pf*}{{\bf Definition.}}
Let $\lbr {\bf C}_{\innu},{\bf S}_{\innu},{\bf C}_{\innv},{\bf S}_{\innv} \rbr$ denote four Hermitian operators written in terms of simple combinations associated 
with ${\bf U}$ and ${\bf V}$, \ie, ${\bf C}_{\innu} \coloneq \mathrm{Re} [{\bf U}]$, ${\bf S}_{\innu} \coloneq \mathrm{Im} [{\bf U}]$, ${\bf C}_{\innv} \coloneq 
\mathrm{Re} [{\bf V}]$, and ${\bf S}_{\innv} \coloneq \mathrm{Im} [{\bf V}]$. The commutation relations involving these operators exhibit a direct connection with 
certain anticommutation relations:
\brr
& & \lbk {\bf C}_{\innu},{\bf C}_{\innv} \rbk = \mathrm{i} A \lbr {\bf S}_{\innu},{\bf S}_{\innv} \rbr , \quad \lbk {\bf C}_{\innu},{\bf S}_{\innv} \rbk = - 
\mathrm{i} A \lbr {\bf S}_{\innu},{\bf C}_{\innv} \rbr , \nn \\
& & \lbk {\bf S}_{\innu},{\bf S}_{\innv} \rbk = \mathrm{i} A \lbr {\bf C}_{\innu},{\bf C}_{\innv} \rbr , \quad \lbk {\bf S}_{\innu},{\bf C}_{\innv} \rbk = - 
\mathrm{i} A \lbr {\bf C}_{\innu},{\bf S}_{\innv} \rbr , \nn 
\err
where the parameter $A$ was previously defined in Eq. (\ref{s2e1}). These results lead us, in principle, to conclude that partial information on two particular 
elements of the set is not complete, since the complementary elements are also necessary to fully characterize the commutator $[ {\bf U},{\bf V} ]$. In this sense, 
let us now consider the four RS uncertainty principles below:
\brr
\lb{s3e6}
\mathscr{V}_{\inc_{\innu}} \mathscr{V}_{\inc_{\innv}} &\geq& \mathscr{C}^{2}({\bf C}_{\innu},{\bf C}_{\innv}) + \frac{1}{4} \left| \lgg \lbk 
{\bf C}_{\innu},{\bf C}_{\innv} \rbk \rgg \right|^{2} , \\
\lb{s3e7}
\mathscr{V}_{\inc_{\innu}} \mathscr{V}_{\ins_{\innv}} &\geq& \mathscr{C}^{2}({\bf C}_{\innu},{\bf S}_{\innv}) + \frac{1}{4} \left| \lgg \lbk 
{\bf C}_{\innu},{\bf S}_{\innv} \rbk \rgg \right|^{2} , \\
\lb{s3e8}
\mathscr{V}_{\ins_{\innu}} \mathscr{V}_{\inc_{\innv}} &\geq& \mathscr{C}^{2}({\bf S}_{\innu},{\bf C}_{\innv}) + \frac{1}{4} \left| \lgg \lbk 
{\bf S}_{\innu},{\bf C}_{\innv} \rbk \rgg \right|^{2} , \\
\lb{s3e9}
\mathscr{V}_{\ins_{\innu}} \mathscr{V}_{\ins_{\innv}} &\geq& \mathscr{C}^{2}({\bf S}_{\innu},{\bf S}_{\innv}) + \frac{1}{4} \left| \lgg \lbk 
{\bf S}_{\innu},{\bf S}_{\innv} \rbk \rgg \right|^{2} .
\err
In addition, the extra result ${\bf C}_{\innu (\innv)}^{2} + {\bf S}_{\innu (\innv)}^{2} = {\bf 1}$ resembles a well-known mathematical property associated with 
the trigonometric functions cosine and sine. For this reason, these Hermitian operators will be henceforth termed `cosine' and `sine' operators, whose respective 
variances can be shown to obey the mathematical identity 
\bd
( \mathscr{V}_{\inc_{\innu}} + \mathscr{V}_{\ins_{\innu}} )( \mathscr{V}_{\inc_{\innv}} + \mathscr{V}_{\ins_{\innv}} ) = \mathscr{V}_{\innu} \mathscr{V}_{\innv} . 
\ed
To make complete this definition, it is interesting to observe that certain combinations of $\lgg {\bf C}_{\innu (\innv)} \rgg^{2}$ and $\lgg {\bf S}_{\innu (\innv)}
\rgg^{2}$ also yield the additional result 
\bd
\lpar \lgg {\bf C}_{\innu} \rgg^{2} + \lgg {\bf S}_{\innu} \rgg^{2} \rpar \lpar \lgg {\bf C}_{\innv} \rgg^{2} + \lgg {\bf S}_{\innv} \rgg^{2} \rpar =
( 1 - \mathscr{V}_{\innu} ) ( 1 - \mathscr{V}_{\innv} ) , 
\ed
which proves itself useful in our subsequent calculations. 
\end{pf*}

Next, by means of mathematical remarks, we establish an important set of other useful results for the sine and cosine operators, whose relevance 
is intrinsically connected with the hierarchy relations involving the uncertainty principles related to ${\bf U}$ and ${\bf V}$ which generalize the 
Massar-Spindel inequality. It is worth mentioning that some proofs demand a logical sequence of algebraic manipulations to be detailed in the text.

\begin{pf*}{{\bf Remark 1.}}
Although the sine and cosine operators are genuinely defined for unitary operators, we shall employ, in due time, the same definition for a general operator
${\bf X}$ as well. In this general case, it can be easily demonstrated that the following properties hold: $\lgg {\bf C}_{\innx}^{2} \rgg + \lgg {\bf S}_{\innx}^{2} 
\rgg = \lgg \half \{ {\bf X},{\bf X}^{\dagger} \} \rgg$, $\lgg {\bf C}_{\innx} \rgg^{2} + \lgg {\bf S}_{\innx} \rgg^{2} = | \lgg {\bf X} \rgg |^{2}$, and
$\mathscr{V}_{\inc_{\innx}} + \mathscr{V}_{\ins_{\innx}} = \mathscr{C}({\bf X},{\bf X}^{\dagger})$. Note that for a normal operator ${\bf N}$ (which satisfies 
$[ {\bf N},{\bf N}^{\dagger} ] = 0$), the covariance function between ${\bf N}$ and ${\bf N}^{\dagger}$ matches the variance of ${\bf N}$, namely,
$\mathscr{V}_{\inc_{\inn}} + \mathscr{V}_{\ins_{\inn}} = \mathscr{V}_{\inn}$.
\end{pf*}

\begin{pf*}{{\bf Remark 2.}}
Let us initially consider the sum of all aforementioned RS uncertainty principles for the sine and cosine operators, as well as the connection between 
the commutation and anticommuation relations of such Hermitian operators. Adequate algebraic manipulations allow, in principle, to obtain an inequality 
for the product $\mathscr{V}_{\innu} \mathscr{V}_{\innv}$ with remarkable mathematical features, \ie,
\be
\lb{s3e10}
( 1 + A^{2} ) \mathscr{V}_{\innu} \mathscr{V}_{\innv} \geq ( 1 + A^{2} ) \mathcal{F}({\bf U},{\bf V}) + A^{2} \, \mathcal{H}({\bf U},{\bf V}) 
\ee
where
\brr
\mathcal{F}({\bf U},{\bf V}) &=& \mathscr{C}^{2}({\bf C}_{\innu},{\bf C}_{\innv}) + \mathscr{C}^{2}({\bf C}_{\innu},{\bf S}_{\innv}) +
\mathscr{C}^{2}({\bf S}_{\innu},{\bf C}_{\innv}) + \mathscr{C}^{2}({\bf S}_{\innu},{\bf S}_{\innv}) , \nn \\
\mathcal{H}({\bf U},{\bf V}) &=& \left| \lgg {\bf C}_{\innu} {\bf C}_{\innv} \rgg \right|^{2} + \left| \lgg {\bf C}_{\innu} {\bf S}_{\innv} \rgg 
\right|^{2} + \left| \lgg {\bf S}_{\innu} {\bf C}_{\innv} \rgg \right|^{2} + \left| \lgg {\bf S}_{\innu} {\bf S}_{\innv} \rgg \right|^{2} . \nn
\err
Note that $\mathcal{F}$ and $\mathcal{H}$ show a nontrivial dependence on the unitary operators ${\bf U}$ and ${\bf V}$, which will be properly discussed 
in the subsequent remarks. Moreover, if compared with Massar-Spindel inequality (\ref{s2e1}), such a result yields subtle additional corrections that will 
depend explicitly on the dimension of the state space. 
\end{pf*}

For completeness sake, let us now establish a first numerical evaluation on the functions $\mathcal{F}$ and $\mathcal{H}$. Figure 2 exhibits the plots of $\mathcal{F}$ 
versus $\mathcal{H}$ for (a) $N=4$ and (b) $N=5$ with the same number of normalized random states used in the previous figure. Since the solid line represents 
$\mathcal{F} = \mathcal{H}$ in both the situations, this preliminary numerical search demonstrates a greater contribution coming from $\mathcal{H}$ than from
$\mathcal{F}$ for most states of low dimension $N$. This fact appeals for a detailed formal investigation on the origins of such contributions, in which each term 
of $\mathcal{F}$ and $\mathcal{H}$ would be examined separately. 
\begin{figure}[!t]
\centering
\begin{minipage}[b]{0.40\linewidth}
\includegraphics[width=\textwidth]{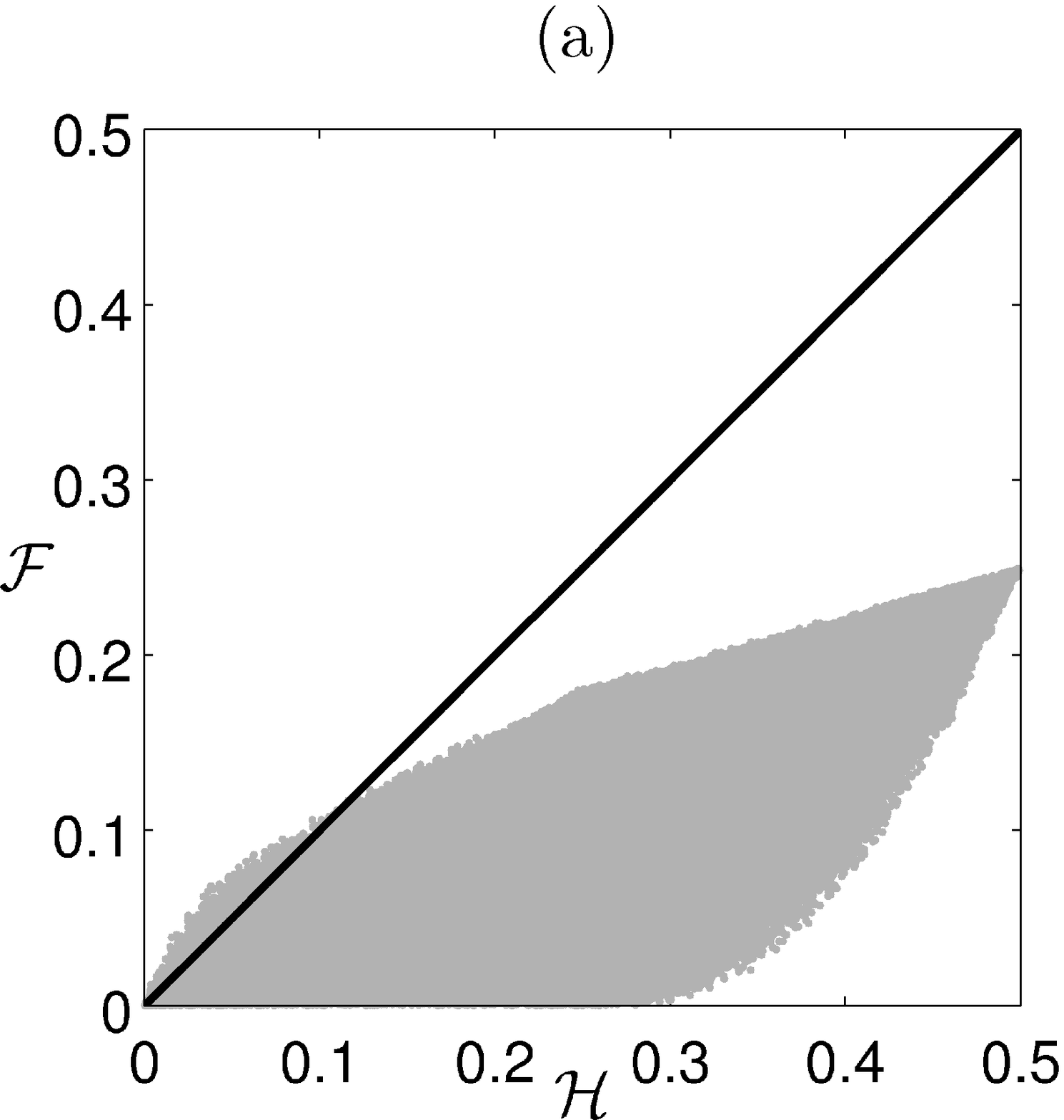}
\end{minipage} \hfill
\begin{minipage}[b]{0.40\linewidth}
\includegraphics[width=\textwidth]{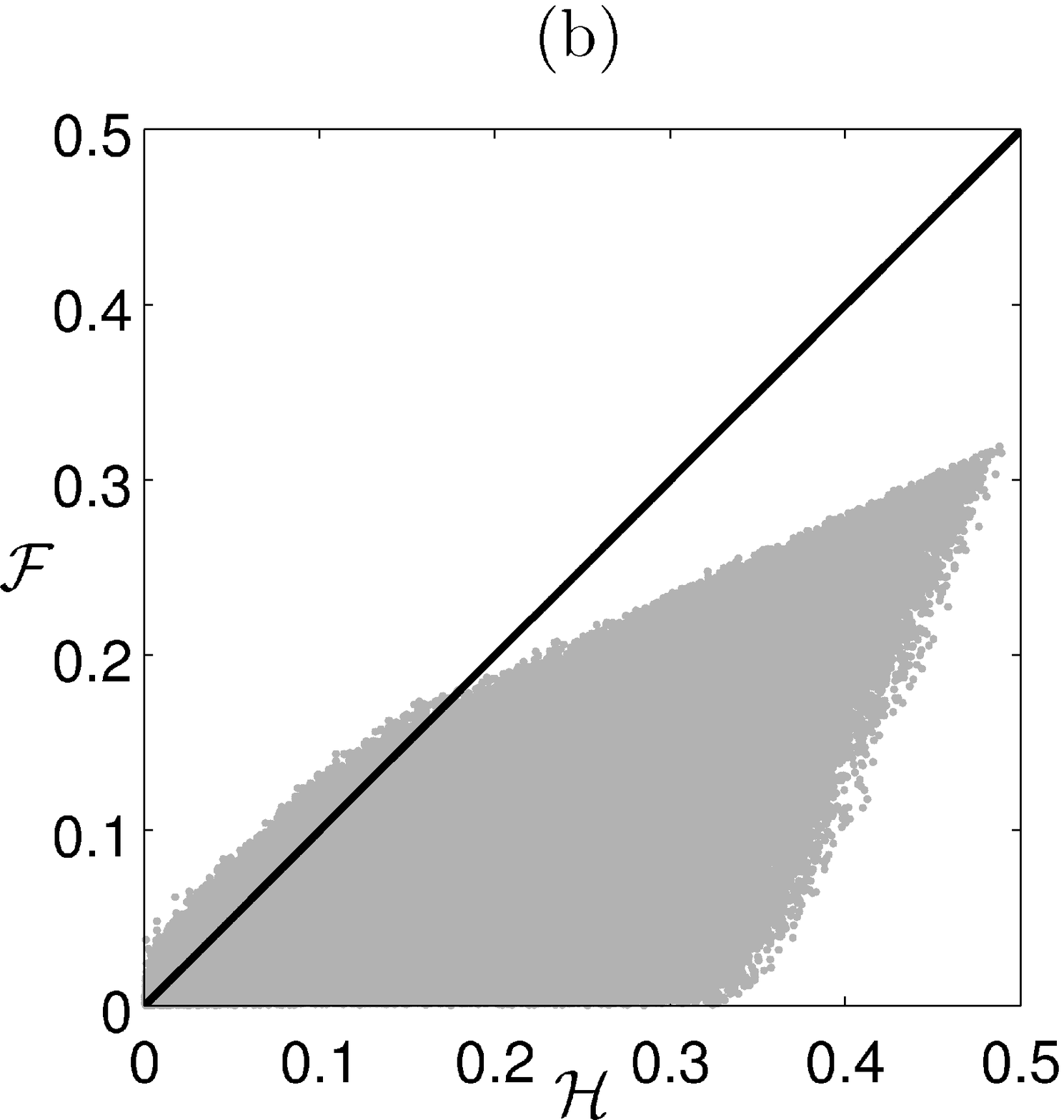}
\end{minipage} 
\caption{Plots of $\mathcal{F}({\bf U},{\bf V})$ versus $\mathcal{H}({\bf U},{\bf V})$ with $10^{6}$ states of randomly generated $\ro$ for two 
different values of dimension $N$, namely, (a) $N=4$ and (b) $N=5$. The solid line corresponds to the case $\mathcal{F} = \mathcal{H}$ in both the 
pictures, where the limit situation $\mathcal{F} = \mathcal{H} = 0$ describes the localized bases. Our numerical computations demonstrate that, in 
principle, the contributions due to $\mathcal{H}$ are more significant than those coming from $\mathcal{F}$ for a large number of random states.} 
\end{figure}

\begin{pf*}{{\bf Remark 3.}}
Important mathematical properties of $\mathcal{H}({\bf U},{\bf V})$ are easily attained examining each of its terms separately: for instance, to calculate 
the modulus squared of $\lgg {\bf C}_{\innu} {\bf C}_{\innv} \rgg$, we initially expand the cosine operators ${\bf C}_{\innu}$ and ${\bf C}_{\innv}$ in terms 
of the Schwinger unitary operators ${\bf U}$ and ${\bf V}$; the next step consists in decomposing the resulting terms into their real and imaginary Hermitian 
parts, whose final expression assumes the form
\brr
\lgg {\bf C}_{\innu} {\bf C}_{\innv} \rgg &=& \frac{1}{4} \lpar \lgg {\bf C}_{\innu \innv} \rgg + \lgg {\bf C}_{\innu \innv^{\dagger}} \rgg +
\lgg {\bf C}_{\innu^{\dagger} \innv} \rgg + \lgg {\bf C}_{\innu^{\dagger} \innv^{\dagger}} \rgg \rpar \nn \\
& & + \, \frac{\mathrm{i}}{4} \lpar \lgg {\bf S}_{\innu \innv} \rgg + \lgg {\bf S}_{\innu \innv^{\dagger}} \rgg + \lgg {\bf S}_{\innu^{\dagger} \innv} \rgg + 
\lgg {\bf S}_{\innu^{\dagger} \innv^{\dagger}} \rgg \rpar . \nn
\err
Once the remaining terms are obtained in the same way, some algebraic simplification gives rise to the following expression for $\mathcal{H}$: 
\brr
\mathcal{H}({\bf U},{\bf V}) &=& \frac{1}{4} \lpar \lgg {\bf C}_{\innu \innv} \rgg^{2} + \lgg {\bf C}_{\innu \innv^{\dagger}} \rgg^{2} +
\lgg {\bf C}_{\innu^{\dagger} \innv} \rgg^{2} + \lgg {\bf C}_{\innu^{\dagger} \innv^{\dagger}} \rgg^{2} \rpar \nn \\
& & + \, \frac{1}{4} \lpar \lgg {\bf S}_{\innu \innv} \rgg^{2} + \lgg {\bf S}_{\innu \innv^{\dagger}} \rgg^{2} + \lgg {\bf S}_{\innu^{\dagger} \innv} \rgg^{2} + 
\lgg {\bf S}_{\innu^{\dagger} \innv^{\dagger}} \rgg^{2} \rpar . \nn
\err
However, this result does not represent a convenient form of $\mathcal{H}$ since the proper summation of $\lgg {\bf C}_{\inno} \rgg^{2} + \lgg {\bf S}_{\inno} 
\rgg^{2} = 1 - \mathscr{V}_{\inno}$ with $\lgg {\bf C}_{\inno}^{2} \rgg + \lgg {\bf S}_{\inno}^{2} \rgg = 1$ (in this situation, one considers all possible 
matches of $\non \equiv {\bf U} {\bf V}, {\bf U} {\bf V}^{\dagger}, {\bf U}^{\dagger} {\bf V}, {\bf U}^{\dagger} {\bf V}^{\dagger}$) permits us to derive a 
simplified expression for such a function that includes the variances $\mathscr{V}_{\innu \innv}$ and $\mathscr{V}_{\innu \innv^{\dagger}}$. Indeed, since 
$\mathscr{V}_{\innu^{\dagger} \innv} \equiv \mathscr{V}_{\innu \innv^{\dagger}}$ and $\mathscr{V}_{\innu^{\dagger} \innv^{\dagger}} \equiv \mathscr{V}_{\innu \innv}$,
it turns immediate to prove that
\be
\lb{s3e11}
\mathcal{H}({\bf U},{\bf V}) = 1 - \frac{1}{2} \lpar \mathscr{V}_{\innu \innv} + \mathscr{V}_{\innu \innv^{\dagger}} \rpar = | \lgg {\bf U} {\bf V} \rgg |^{2} + 
| \lgg {\bf U} {\bf V}^{\dagger} \rgg |^{2}
\ee
which is invariant under the transformations ${\bf U} \rightarrow \mathrm{e}^{\mathrm{i} \varphi} {\bf U}$ and ${\bf V} \rightarrow \mathrm{e}^{\mathrm{i} \theta} {\bf V}$
for $\{ \varphi,\theta \} \in \mathbb{R}$.
\end{pf*} 

\begin{pf*}{{\bf Remark 4.}}
Note that Eq. (\ref{s3e11}) represents a lower bound for any element of\footnote{Let $\left\| {\bf U} \right\|_{\mathrm{HS}} = \left\| {\bf V} \right\|_{\mathrm{HS}} = 
\sqrt{N}$ and $\left\| {\bf C}_{\innu (\innv)} \right\|_{\mathrm{HS}} = \left\| {\bf S}_{\innu (\innv)} \right\|_{\mathrm{HS}} = \sqrt{\frac{N}{2}}$ characterize the 
Hilbert-Schmidt norms associated with the Schwinger unitary operators and their respective related trigonometric operators, where $\left\| {\bf X} \right\|_{\mathrm{HS}} 
\coloneq \sqrt{\mathrm{Tr} \lbk {\bf X}^{\dagger} {\bf X} \rbk}$ defines the aforementioned norm \cite{Bhatia}. The further mathematical property
\bd
\Bigl\{ \left\| {\bf C}_{\innu} {\bf C}_{\innv} \right\|_{\mathrm{HS}}, \left\| {\bf C}_{\innu} {\bf S}_{\innv} \right\|_{\mathrm{HS}},
\left\| {\bf S}_{\innu} {\bf C}_{\innv} \right\|_{\mathrm{HS}}, \left\| {\bf S}_{\innu} {\bf S}_{\innv} \right\|_{\mathrm{HS}} \Bigr\} \leq \frac{N}{2}
\ed
represents an effective gain in this stage, since it brings out relevant information on the different products of cosine and sine operators used in the text.}  
\bd
\lbr | \lgg {\bf C}_{\innu} {\bf C}_{\innv} \rgg |^{2}, | \lgg {\bf C}_{\innu} {\bf S}_{\innv} \rgg |^{2}, | \lgg {\bf S}_{\innu} {\bf C}_{\innv} \rgg |^{2},
| \lgg {\bf S}_{\innu} {\bf S}_{\innv} \rgg |^{2} \rbr \geq \frac{1}{4} \mathcal{H}({\bf U},{\bf V}) ,
\ed
this fact being discussed by Massar and Spindel \cite{Massar} through different mathematical arguments. In fact, the authors demonstrated that for 
a given choice of phases of the Schwinger unitary operators, the restrictions $\lgg {\bf C}_{\innu (\innv)} \rgg \in \mathbb{R}_{+} \Rightarrow \lgg 
{\bf S}_{\innu (\innv)} \rgg = 0$ select a set $\{ \ro \}$ of finite quantum states for which the inequality 
\bd
| \lgg {\bf C}_{\innu} {\bf C}_{\innv} \rgg | \geq \sqrt{( 1 - \mathscr{V}_{\innu} )( 1 - \mathscr{V}_{\innv} )} - \sqrt{\mathscr{V}_{\innu} 
\mathscr{V}_{\innv}} 
\ed
can be formally verified and also numerically tested. Despite the correlations between ${\bf U}$ and ${\bf V}$ do not appear in such an expression, the comparison
with $\mathcal{H}$ is unavoidable in this case, since correlations represent an important quantum effect that deserve our attention. The saturation is reached in 
both the situations for localized bases.
\end{pf*}

\begin{pf*}{{\bf Remark 5.}}
Let us now decompose $\mathcal{F}$ into three terms $\mathcal{F}_{1}$, $\mathcal{F}_{2}$ and $\mathcal{F}_{3}$, whose different contributions will be formally 
calculated with the help of the mathematical procedure sketched in the previous remarks. We initially consider the term
\bd
\mathcal{F}_{1} = \frac{1}{4} \lpar \lgg \{ {\bf C}_{\innu},{\bf C}_{\innv} \} \rgg^{2} + \lgg \{ {\bf C}_{\innu},{\bf S}_{\innv} \} \rgg^{2} + 
\lgg \{ {\bf S}_{\innu},{\bf C}_{\innv} \} \rgg^{2} + \lgg \{ {\bf S}_{\innu},{\bf S}_{\innv} \} \rgg^{2} \rpar
\ed
responsible for contributions associated with squared mean values of all anticommutation relations involved in $\mathcal{F}$ through the covariance
functions. So, let $\lgg \{ {\bf C}_{\innu},{\bf C}_{\innv} \} \rgg$ admit the form
\brr
\lgg \{ {\bf C}_{\innu},{\bf C}_{\innv} \} \rgg &=& \frac{1}{4} \lbk ( 1 + \om ) \lpar \lgg {\bf C}_{\innu \innv} \rgg + \lgg
{\bf C}_{\innu^{\dagger} \innv^{\dagger}} \rgg \rpar + ( 1 + \om^{\ast} ) \lpar \lgg {\bf C}_{\innu \innv^{\dagger}} \rgg + \lgg
{\bf C}_{\innu^{\dagger} \innv} \rgg \rpar \rbk \nn \\
& & + \frac{\mathrm{i}}{4} \lbk ( 1 + \om ) \lpar \lgg {\bf S}_{\innu \innv} \rgg + \lgg {\bf S}_{\innu^{\dagger} \innv^{\dagger}} \rgg \rpar + 
( 1 + \om^{\ast} ) \lpar \lgg {\bf S}_{\innu \innv^{\dagger}} \rgg + \lgg {\bf S}_{\innu^{\dagger} \innv} \rgg \rpar \rbk \nn
\err
in complete analogy with $\lgg {\bf C}_{\innu} {\bf C}_{\innv} \rgg$. Writing the remaining mean values in an analogous way, subsequent computations allow 
to prove that $\mathcal{F}_{1}$ is connected with $\mathcal{H}({\bf U},{\bf V})$ through the relation $\mathcal{F}_{1} = \cos^{2} \lpar \frac{\Phi}{2} \rpar 
\mathcal{H}$ for $\Phi = \frac{2 \pi}{N}$ fixed. 

Following, in what concerns the second term 
\brr
\mathcal{F}_{2} &=& \lgg {\bf C}_{\innu} \rgg \lgg {\bf C}_{\innv} \rgg \lgg \{ {\bf C}_{\innu},{\bf C}_{\innv} \} \rgg + \lgg {\bf C}_{\innu} \rgg \lgg 
{\bf S}_{\innv} \rgg \lgg \{ {\bf C}_{\innu},{\bf S}_{\innv} \} \rgg \nn \\
& & + \, \lgg {\bf S}_{\innu} \rgg \lgg {\bf C}_{\innv} \rgg \lgg \{ {\bf S}_{\innu},{\bf C}_{\innv} \} \rgg + \lgg {\bf S}_{\innu} \rgg \lgg {\bf S}_{\innv} 
\rgg \lgg \{ {\bf S}_{\innu},{\bf S}_{\innv} \} \rgg , \nn
\err
it can be expressed by means of the convenient form
\brr
2 \mathcal{F}_{2} &=& \mathrm{Re} [ \lgg {\bf U} \rgg \lgg {\bf V} \rgg ] \mathrm{Re} [ ( 1 + \om ) \lgg {\bf U} {\bf V} \rgg ] + \mathrm{Re} [ 
\lgg {\bf U} \rgg \lgg {\bf V}^{\dagger} \rgg ] \mathrm{Re} [ ( 1 + \om^{\ast} ) \lgg {\bf U} {\bf V}^{\dagger} \rgg ] \nn \\
& & + \mathrm{Im} [ \lgg {\bf U} \rgg \lgg {\bf V} \rgg ] \mathrm{Im} [ ( 1 + \om ) \lgg {\bf U} {\bf V} \rgg ] + \mathrm{Im} [ \lgg {\bf U} \rgg \lgg 
{\bf V}^{\dagger} \rgg ] \mathrm{Im} [ ( 1 + \om^{\ast} ) \lgg {\bf U} {\bf V}^{\dagger} \rgg ] \nn
\err
which represents an important formal result in our calculations. 

Finally, let us briefly mention that $\mathcal{F}_{3}$ coincides with $( 1 - \mathscr{V}_{\innu} ) ( 1 - \mathscr{V}_{\innv} )$ since it is equivalent to the 
product $\lpar \lgg {\bf C}_{\innu} \rgg^{2} + \lgg {\bf S}_{\innu} \rgg^{2} \rpar \lpar \lgg {\bf C}_{\innv} \rgg^{2} + \lgg {\bf S}_{\innv} \rgg^{2} \rpar$;
consequently, the function
\be
\lb{s3e12} 
\mathcal{F}({\bf U},{\bf V}) = \mathcal{F}_{1}({\bf U},{\bf V}) - \mathcal{F}_{2}({\bf U},{\bf V}) + \mathcal{F}_{3}({\bf U},{\bf V})
\ee
can be immediately obtained. Note that the nontrivial dependence of Eq. (\ref{s3e10}) on the Schwinger unitary operators is completely justified in these remarks.
\end{pf*}

This set of mathematical remarks establishes a first solid algebraic framework for the unitary operators ${\bf U}$ and ${\bf V}$, whose intrinsic virtues
lead us to formulate a theorem which generalizes the Massar-Spindel inequality (\ref{s2e1}) in order to include the quantum correlation effects between the 
aforementioned operators. In fact, this theorem consists of an initial compilation of efforts in our future search for the tightest bound, where the 
correlation function has occupied an important place in the investigative process.

\s{\begin{pf*}{{\bf Theorem.}}
{\it Let ${\bf U}$ and ${\bf V}$ be two unitary operators defined in a $N$-dimensional state vector space which satisfy the commutation relations 
$\lbk {\bf U},{\bf V} \rbk = (1-\om) {\bf U} {\bf V}$ and $\lbk {\bf U},{\bf V}^{\dagger} \rbk = (1-\om^{\ast}) {\bf U} {\bf V}^{\dagger}$ for 
$\om \coloneq \exp \lpar \frac{2 \pi \mathrm{i}}{N} \rpar$. Moreover, let $\mathscr{V}_{\innu} \coloneq 1 - | \lgg {\bf U} \rgg |^{2}$ and 
$\mathscr{V}_{\innv} \coloneq 1 - | \lgg {\bf V} \rgg |^{2}$ represent the respective variances such that $0 \leq \mathscr{V}_{\innu (\innv)} \leq 1$. 
The inequality 
\be
\lb{s3e13}
\mathscr{V}_{\innu} \mathscr{V}_{\innv} \geq \mathcal{F}({\bf U},{\bf V}) + \sin^{2} \lpar \frac{\pi}{N} \rpar \mathcal{H}({\bf U},{\bf V})
\ee
yields a new bound for $\mathscr{V}_{\innu} \mathscr{V}_{\innv}$, where the quantum correlation effects related to the unitary operators are taken 
into account. Note that $\mathcal{F}$ and $\mathcal{H}$ were precisely defined and formally studied in the previous mathematical remarks, the saturation 
$\mathcal{F} = \mathcal{H} = 0$ being attained for localized bases.}
\end{pf*}}

Next, let us determine some additional results focused on the Hermitian operators $\lbr {\bf C}_{\delta \innu},{\bf S}_{\delta \innu},{\bf C}_{\delta \innv},
{\bf S}_{\delta \innv} \rbr$ in order to yield a set of specific intermediate inequalities whose hierarchical relations correspond to a solid bridge towards 
the tightest bound. 

\subsection{Hierarchical relations}

We start this subsection stating a first important result for the sine and cosine operators $\lbr {\bf C}_{\delta \innu},{\bf S}_{\delta \innu},
{\bf C}_{\delta \innv},{\bf S}_{\delta \innv} \rbr$ defined, in turn, in terms of the non-unitary counterparts $\delta {\bf U}$ and $\delta {\bf V}$. 
This particular result shows how the mean values of their commutation and anticommutation relations are linked with determined correlation functions, namely,
\brr
& & \lgg [ {\bf C}_{\delta \innu},{\bf C}_{\delta \innv} ] \rgg = \mathrm{i} A \, \lgg \{ {\bf S}_{\delta \innu},{\bf S}_{\delta \innv} \} \rgg = 2 \mathrm{i} A \,
\mathscr{C}({\bf S}_{\delta \innu},{\bf S}_{\delta \innv}) , \nn \\
& & \lgg [ {\bf C}_{\delta \innu},{\bf S}_{\delta \innv} ] \rgg = - \mathrm{i} A \, \lgg \{ {\bf S}_{\delta \innu},{\bf C}_{\delta \innv} \} \rgg = - 2 \mathrm{i} A \,
\mathscr{C}({\bf S}_{\delta \innu},{\bf C}_{\delta \innv}) , \nn \\
& & \lgg [ {\bf S}_{\delta \innu},{\bf C}_{\delta \innv} ] \rgg = - \mathrm{i} A \, \lgg \{ {\bf C}_{\delta \innu},{\bf S}_{\delta \innv} \} \rgg = - 2 \mathrm{i} A \,
\mathscr{C}({\bf C}_{\delta \innu},{\bf S}_{\delta \innv}) , \nn \\
& & \lgg [ {\bf S}_{\delta \innu},{\bf S}_{\delta \innv} ] \rgg = \mathrm{i} A \lpar \lgg \{ {\bf C}_{\delta \innu},{\bf C}_{\delta \innv} \} \rgg + 2 \rpar = 
2 \mathrm{i} A \lbk \mathscr{C}({\bf C}_{\delta \innu},{\bf C}_{\delta \innv}) + 1 \rbk . \nn
\err
Following, let us complete this set of results with relations that connect all the covariance functions previously defined with the original framework:
\brr
\mathcal{F}_{3} \, \mathscr{C}({\bf C}_{\delta \innu},{\bf C}_{\delta \innv}) &=& \mathscr{C}({\bf C}_{\innu},{\bf C}_{\innv}) 
\lgg {\bf C}_{\innu} \rgg \lgg {\bf C}_{\innv} \rgg + \mathscr{C}({\bf S}_{\innu},{\bf S}_{\innv}) \lgg {\bf S}_{\innu} \rgg \lgg {\bf S}_{\innv} \rgg \nn \\
& & + \mathscr{C}({\bf C}_{\innu},{\bf S}_{\innv}) \lgg {\bf C}_{\innu} \rgg \lgg {\bf S}_{\innv} \rgg + \mathscr{C}({\bf S}_{\innu},{\bf C}_{\innv})
\lgg {\bf S}_{\innu} \rgg \lgg {\bf C}_{\innv} \rgg , \nn \\
\mathcal{F}_{3} \, \mathscr{C}({\bf C}_{\delta \innu},{\bf S}_{\delta \innv}) &=& - \mathscr{C}({\bf C}_{\innu},{\bf C}_{\innv}) 
\lgg {\bf C}_{\innu} \rgg \lgg {\bf S}_{\innv} \rgg + \mathscr{C}({\bf S}_{\innu},{\bf S}_{\innv}) \lgg {\bf S}_{\innu} \rgg \lgg {\bf C}_{\innv} \rgg \nn \\
& & + \mathscr{C}({\bf C}_{\innu},{\bf S}_{\innv}) \lgg {\bf C}_{\innu} \rgg \lgg {\bf C}_{\innv} \rgg - \mathscr{C}({\bf S}_{\innu},{\bf C}_{\innv})
\lgg {\bf S}_{\innu} \rgg \lgg {\bf S}_{\innv} \rgg , \nn \\
\mathcal{F}_{3} \, \mathscr{C}({\bf S}_{\delta \innu},{\bf C}_{\delta \innv}) &=& - \mathscr{C}({\bf C}_{\innu},{\bf C}_{\innv}) 
\lgg {\bf S}_{\innu} \rgg \lgg {\bf C}_{\innv} \rgg + \mathscr{C}({\bf S}_{\innu},{\bf S}_{\innv}) \lgg {\bf C}_{\innu} \rgg \lgg {\bf S}_{\innv} \rgg \nn \\
& & - \mathscr{C}({\bf C}_{\innu},{\bf S}_{\innv}) \lgg {\bf S}_{\innu} \rgg \lgg {\bf S}_{\innv} \rgg + \mathscr{C}({\bf S}_{\innu},{\bf C}_{\innv})
\lgg {\bf C}_{\innu} \rgg \lgg {\bf C}_{\innv} \rgg , \nn \\
\mathcal{F}_{3} \, \mathscr{C}({\bf S}_{\delta \innu},{\bf S}_{\delta \innv}) &=& \mathscr{C}({\bf C}_{\innu},{\bf C}_{\innv}) 
\lgg {\bf S}_{\innu} \rgg \lgg {\bf S}_{\innv} \rgg + \mathscr{C}({\bf S}_{\innu},{\bf S}_{\innv}) \lgg {\bf C}_{\innu} \rgg \lgg {\bf C}_{\innv} \rgg \nn \\
& & - \mathscr{C}({\bf C}_{\innu},{\bf S}_{\innv}) \lgg {\bf S}_{\innu} \rgg \lgg {\bf C}_{\innv} \rgg - \mathscr{C}({\bf S}_{\innu},{\bf C}_{\innv})
\lgg {\bf C}_{\innu} \rgg \lgg {\bf S}_{\innv} \rgg . \nn
\err
These equations enable us to rewrite $\mathcal{F}$ in the compact form
\be
\lb{s3e14}
\! \frac{\mathcal{F}}{\mathcal{F}_{3}} = \mathscr{C}^{2}({\bf C}_{\delta \innu},{\bf C}_{\delta \innv}) + \mathscr{C}^{2}({\bf C}_{\delta \innu},
{\bf S}_{\delta \innv}) + \mathscr{C}^{2}({\bf S}_{\delta \innu},{\bf C}_{\delta \innv}) + \mathscr{C}^{2}({\bf S}_{\delta \innu},{\bf S}_{\delta \innv}) ,
\ee
while $( \mathscr{V}_{\inc_{\delta \innu}} + \mathscr{V}_{\ins_{\delta \innu}} ) ( \mathscr{V}_{\inc_{\delta \innv}} + \mathscr{V}_{\ins_{\delta \innv}} ) = 
\mathscr{V}_{\delta \innu} \mathscr{V}_{\delta \innv}$ brings out a completely analogous expression to that established for $\mathscr{V}_{\innu} 
\mathscr{V}_{\innv}$; consequently,
\be
\lb{s3e15}
\mathscr{V}_{\delta \innu} \mathscr{V}_{\delta \innv} \geq \frac{\mathcal{F}}{\mathcal{F}_{3}} + \sin^{2} \lpar \frac{\pi}{N} \rpar 
\frac{\mathcal{H}}{\mathcal{F}_{3}}
\ee
represents an alternative form of Eq. (\ref{s3e13}) since $\mathscr{V}_{\innu} \mathscr{V}_{\innv} \equiv \mathcal{F}_{3} \mathscr{V}_{\delta \innu}
\mathscr{V}_{\delta \innv}$. 

Now, let us rewrite Eq. (\ref{s3e15}) according to definitions employed for $\mathcal{S}_{\delta \innu}$ and $\mathcal{S}_{\delta \innv}$ in the previous section,
\be
\lb{s3e16}
\mathcal{S}_{\delta \innu} \mathcal{S}_{\delta \innv} \geq \lbk 1 + \sin \lpar \frac{2 \pi}{N} \rpar \rbk \!\! \lbk \csc^{2} \lpar \frac{\pi}{N} \rpar
\frac{\mathcal{F}}{\mathcal{F}_{3}} + \frac{\mathcal{H}}{\mathcal{F}_{3}} \rbk .
\ee
From this result, it is relatively easy to derive a number of simpler (but looser) inequalities: for instance, given the non-negativity of
$\frac{\mathcal{F}}{\mathcal{F}_{3}}$ [cf. Eq. (\ref{s3e14})], the first term in the second bracket can be dropped to give
\bd
\mathcal{S}_{\delta \innu} \mathcal{S}_{\delta \innv} \geq \lbk 1 + \sin \lpar \frac{2 \pi}{N} \rpar \rbk \frac{\mathcal{H}}{\mathcal{F}_{3}} ,
\ed
which corresponds, in such a case, to the HKR uncertainty principle; alternatively, $\frac{\mathcal{H}}{\mathcal{F}_{3}}$ can also be dropped (for the same 
reason, \ie, its non-negativity) in order to yield 
\bd
\mathcal{S}_{\delta \innu} \mathcal{S}_{\delta \innv} \geq \lbk 1 + \sin \lpar \frac{2 \pi}{N} \rpar \rbk \csc^{2} \lpar \frac{\pi}{N} \rpar
\frac{\mathcal{F}}{\mathcal{F}_{3}} .
\ed
Consequently, such inequalities are then combined in order to give a tighter one,
\be
\lb{s3e17}
\mathcal{S}_{\delta \innu} \mathcal{S}_{\delta \innv} \geq \lbk 1 + \sin \lpar \frac{2 \pi}{N} \rpar \rbk \mbox{max} \lbk \csc^{2} \lpar \frac{\pi}{N} \rpar
\frac{\mathcal{F}}{\mathcal{F}_{3}} , \frac{\mathcal{H}}{\mathcal{F}_{3}} \rbk .
\ee
At this point, one is led to ask if there exists a general ordering relation between $\csc^{2} \lpar \frac{\pi}{N} \rpar \frac{\mathcal{F}}{\mathcal{F}_{3}}$
and $\frac{\mathcal{H}}{\mathcal{F}_{3}}$. To answer this question, let us establish a set of numerical calculations associated with the different values of 
dimension $N$, where, for each specific case, one has approximately $10^{6}$ randomly generated states. In order to make the presentation of these numerical 
results more self-contained, Fig. 3 depicts the plots of $\csc^{2} \lpar \frac{\pi}{N} \rpar \frac{\mathcal{F}}{\mathcal{F}_{3}}$ versus
$\frac{\mathcal{H}}{\mathcal{F}_{3}}$ for $N=2,3,4,6$ and $20$. In this case, note that the maximum can indeed arise from either terms, depending on the particular 
chosen state $\ro$. As a {\sl rule of thumb}, one has the following: for low dimensional states (\eg, $N=2,3$), the term $\frac{\mathcal{H}}{\mathcal{F}_{3}}$ 
usually dominates; however, as the dimension $N$ increases ($N \geq 6$), $\csc^{2} \lpar \frac{\pi}{N} \rpar \frac{\mathcal{F}}{\mathcal{F}_{3}}$ becomes the 
usually dominant term. Finally, if one considers $N=4$, it is visible that there is not a usually dominant term --- in this case, the result of the optimization is 
strongly dependent on the particular input state.
\begin{figure}[!t]
\centering
\begin{minipage}[b]{0.65\linewidth}
\includegraphics[width=\textwidth]{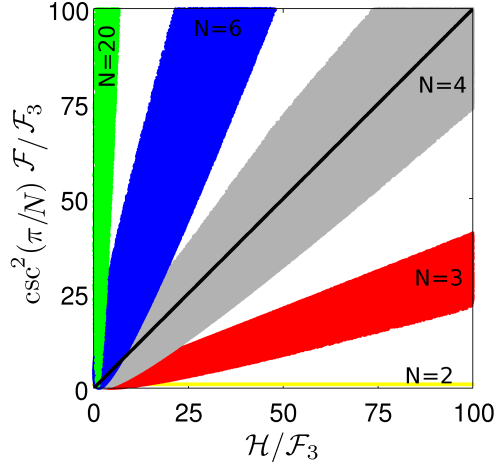}
\end{minipage}
\caption{Plots of $\csc^{2} \lpar \frac{\pi}{N} \rpar \frac{\mathcal{F}}{\mathcal{F}_{3}}$ versus $\frac{\mathcal{H}}{\mathcal{F}_{3}}$ with, at least, 
$10^{6}$ randomly generated states for each different value of dimension $N$. Thus, yellow points depict the $N=2$ case, while red, grey, blue, and green 
points describe, respectively, the $N=3,4,6$, and $20$ situations. Note that $(0,0)$ is associated with localized bases in this picture, whereas the solid 
line corresponds to the particular case $\csc^{2} \lpar \frac{\pi}{N} \rpar \frac{\mathcal{F}}{\mathcal{F}_{3}} = \frac{\mathcal{H}}{\mathcal{F}_{3}}$. It 
is interesting to observe how $\csc^{2} \lpar \frac{\pi}{N} \rpar \frac{\mathcal{F}}{\mathcal{F}_{3}}$ and $\frac{\mathcal{H}}{\mathcal{F}_{3}}$ behave when
different dimensions of state vector space are considered in our numerical computations.} 
\end{figure}

As it turns out, inequality (\ref{s3e17}) can be further strengthened by the addition of yet another (less trivial) lower bound of $\csc^{2} \lpar \frac{\pi}{N} 
\rpar \frac{\mathcal{F}}{\mathcal{F}_{3}} + \frac{\mathcal{H}}{\mathcal{F}_{3}}$ to the maximization at hand. Just as $\frac{\mathcal{F}}{\mathcal{F}_{3}}$ could be 
written as in Eq. (\ref{s3e14}), we can also show that $\frac{\mathcal{H}}{\mathcal{F}_{3}}$ admits the form
\be
\lb{s3e18}
\frac{\mathcal{H}}{\mathcal{F}_{3}} = \sec^{2} \lpar \frac{\pi}{N} \rpar \lbk 1 + 2 \mathscr{C}({\bf C}_{\delta \innu},{\bf C}_{\delta \innv}) +
\frac{\mathcal{F}}{\mathcal{F}_{3}} \rbk .
\ee
With that in mind, some straightforward manipulation yields
\brr
\csc^{2} \lpar \frac{\pi}{N} \rpar \frac{\mathcal{F}}{\mathcal{F}_{3}} + \frac{\mathcal{H}}{\mathcal{F}_{3}} & = & \sec^{2} \lpar \frac{\pi}{N} \rpar
( 1 + x )^{2} + \csc^{2} \lpar \frac{\pi}{N} \rpar x^{2} \nn \\
& & + \, 4 \csc^{2} \lpar \frac{2 \pi}{N} \rpar \lpar \frac{\mathcal{F}}{\mathcal{F}_{3}} - x^{2} \rpar \nn \\
& \geq & \sec^{2} \lpar \frac{\pi}{N} \rpar ( 1 + x )^{2} + \csc^{2} \lpar \frac{\pi}{N} \rpar x^{2} \nn
\err
for $x \equiv \mathscr{C}({\bf C}_{\delta \innu},{\bf C}_{\delta \innv})$, where the inequality follows from fact that $\frac{\mathcal{F}}{\mathcal{F}_{3}} \geq x^{2}$
[cf. Eq. (\ref{s3e14})]. A mathematical virtue of this lower bound is that it depends only on the covariance $\mathscr{C}({\bf C}_{\delta \innu},{\bf C}_{\delta \innv})$, 
whereas $\frac{\mathcal{F}}{\mathcal{F}_{3}}$ and $\frac{\mathcal{H}}{\mathcal{F}_{3}}$ depend on the covariances of all combinations among ${\bf C}_{\delta \innu}$, 
${\bf C}_{\delta \innv}$, ${\bf S}_{\delta \innu}$ and ${\bf S}_{\delta \innv}$. As a result, we may now write
\be
\lb{s3e19}
\mathcal{S}_{\delta \innu} \mathcal{S}_{\delta \innv} \geq \lbk 1 + \sin \lpar \frac{2 \pi}{N} \rpar \rbk \mbox{max} \lbk 1 + \mathcal{G}(x;N) ,  
\csc^{2} \lpar \frac{\pi}{N} \rpar \frac{\mathcal{F}}{\mathcal{F}_{3}} , \frac{\mathcal{H}}{\mathcal{F}_{3}} \rbk
\ee
with $\mathcal{G}(x;N) \coloneq 4 \csc^{2} \lpar \frac{2 \pi}{N} \rpar x^{2} + 2 \sec^{2} \lpar \frac{\pi}{N} \rpar x + \tan^{2} \lpar \frac{\pi}{N} \rpar$. Note 
that $1 + \mathcal{G}(x;N)$ coincides, in such a situation, with $\sec^{2} \lpar \frac{\pi}{N} \rpar (1+x)^{2} + \csc^{2} \lpar \frac{\pi}{N} \rpar x^{2}$. Once again, 
numerical calculations demonstrate that there is not a general ordering between the arguments of the maximization of such an equation.

It is worth stressing that both bounds of Eqs. (\ref{s3e16}) and (\ref{s3e19}) explicitly depend on the Hilbert space dimension, as well as on the particular state 
under consideration. Henceforth, we further loose the bound (\ref{s3e19}) to provide yet another bound, but now a state-independent one (\ie, solely dependent on the
Hilbert space dimension). In this way, let us initially consider the inequality
\bd
\mathcal{S}_{\delta \innu} \mathcal{S}_{\delta \innv} \geq \lbk 1 + \sin \lpar \frac{2 \pi}{N} \rpar \rbk \lbk 1 + \mathcal{G}(x;N) \rbk .
\ed
So, for a given $N$, $\mathcal{G}(x;N)$ describes a parabola with upwards concavity and minimum value equal to $0$ ($y$-coordinate of the vertex), namely,
$\mathcal{G}(x;N) \geq 0$, which implies that
\be
\lb{s3e20}
\mathcal{S}_{\delta \innu} \mathcal{S}_{\delta \innv} \geq 1 + \sin \lpar \frac{2 \pi}{N} \rpar . 
\ee
Despite of all the mathematical assumptions used in the relaxation process of inequality (\ref{s3e16}), the bound above is still tighter than the one proposed in 
Ref. \cite{Massar}, which is now trivially proved by disregarding the sine function in the inequality (\ref{s3e20}), that is, $\mathcal{S}_{\delta \innu}
\mathcal{S}_{\delta \innv} \geq 1$.

\h{\begin{pf*}{{\bf Hierarchy.}} 
The tightest bound of the product $\mathcal{S}_{\delta \innu} \mathcal{S}_{\delta \innv}$ is particularly related to an underlying minimization problem --- see Appendix A.
Although this problem seems simple at first glance \cite{Jackiw}, let us now establish a set of inequalities which characterizes an important hierarchical relation among 
certain formal results mentioned in the body of the text: 
\be
\lb{s3e21}
\mathcal{S}_{\delta \innu} \mathcal{S}_{\delta \innv} \geq \mathfrak{R}_{1} \geq \mathfrak{R}_{2} \geq \mathfrak{R}_{3} \geq \mathfrak{R}_{4} \geq 1 , 
\ee
where $\mathfrak{R}_{1} \equiv$ tightest bound, $\mathfrak{R}_{2} =$ RHS of Eq. (\ref{s3e16}), $\mathfrak{R}_{3} = $ RHS of Eq. (\ref{s3e19}), and finally,
$\mathfrak{R}_{4} = $ RHS of Eq. (\ref{s3e20}). Note that $\mathcal{S}_{\delta \innu} \mathcal{S}_{\delta \innv} \geq 1$ describes the loosest bound, which is 
precisely the inequality of Massar and Spindel \cite{Massar}.
\end{pf*}}

\section{The tightest bound}

Massar and Spindel \cite{Massar} relied on the works of Jackiw and Opatrn\'{y} \cite{Jackiw} to provide a numerical recipe that implicitly produces the tightest 
bound for the cloud of points depicted in Fig. 1. In this section, we review such a procedure while applying it to explicitly construct closed-form expressions 
for the tightest bound here related to physical systems characterized by a low-dimensional space of states.

Generally, the tightest bound can be obtained by the following reasoning:
\bd
\!\!\!\! \mathcal{S}_{\delta \innu} \mathcal{S}_{\delta \innv} \geq \min_{| \psi \rgg} \mathcal{S}_{\delta \innu} \mathcal{S}_{\delta \innv} = \eps^{-2} 
\min_{| \psi \rgg} \mathscr{V}_{\delta \innu} \mathscr{V}_{\delta \innv} = \eps^{-2} \mathscr{V}^{(0)}_{\delta \innu} \mathscr{V}^{(0)}_{\delta \innv} =
\lbk \eps^{-1} \mathscr{V}^{(0)}_{\delta \innu} \rbk^{2} = {\mathcal{S}_{\delta \innu}^{(0)}}^{2} .
\ed
In such a case, the super-index ${(0)}$ indicates that the corresponding normalized discrete wavefunction is associated with the nondegenerate ground state 
$| \psi_{0} \rgg$, which minimizes the product $\mathscr{V}_{\innu} \mathscr{V}_{\innv}$. In fact, this ground state will also minimize 
$\mathscr{V}_{\delta \innu} \mathscr{V}_{\delta \innv}$, since $\mathscr{V}_{\delta \mathcal{O}}$ is monotonically increasing with $\mathscr{V}_{\mathcal{O}}$ 
[cf. Eq. (\ref{s2e3})] --- this observation justifies the second equality. Besides, both the variances evaluated with respect to the ground state satisfy
$\mathscr{V}^{(0)}_{\delta \innu} = \mathscr{V}^{(0)}_{\delta \innv}$, which justifies the third equality.

Now, let us establish a sequence of mathematical steps based on the respective eigenvalues and eigenvectors of the Harper Hamiltonian \cite{Barker}
\bd
{\bf H} = - \sin (\th) {\bf C}_{\innu} - \cos (\th) {\bf C}_{\innv}
\ed
for $0 \leq \th \leq \frac{\pi}{2}$. Initially, we look for the smallest eigenvalue of such a Hermitian operator, as well as for its corresponding eigenvector 
in a given fixed $N$-dimensional state vector space. Since the eigenvalues and eigenvectors related to the Harper Hamiltonian are dependent on the angle $\th$, 
let us consider that eigenvector evaluated at the first step in order to estimate the maximum of $\cos (\th) | \lgg {\bf U} \rgg | + \sin (\th) | \lgg {\bf V} \rgg |$ 
for all $\th \in \lbk 0, \frac{\pi}{2} \rbk$ -- which exactly coincides with the smallest eigenvalue in this case. In particular, such a mathematical procedure 
allows us to obtain, through numerical evaluations, the value of $\th = \frac{\pi}{4}$ for any dimension $N$, and also to characterize the ground state 
$| \psi_{0} \rgg \equiv | 0 \rgg_{N}$ with well-defined mathematical properties (see Appendix A for possible connection with Harper functions).

Table 1 illustrates the hierarchical relation depicted in Eq. (\ref{s3e21}) where, in particular, certain numerical results directly related to the analytical 
calculations performed for $\{ \mathfrak{R}_{1},\mathfrak{R}_{2},\mathfrak{R}_{3},\mathfrak{R}_{4} \}$ with $N \in [2,6]$ are exhibited --- closed-form expressions 
for $\mathcal{S}_{\delta U}^{(0)}$ can be viewed in Appendix A and their respective squared values $\mathfrak{R}_{1} \equiv {\mathcal{S}_{\delta \innu}^{(0)}}^{2}$ 
compared with those results previously obtained in Fig. 1. Note that for higher dimensions, some numerical methods (for instance, Laguerre method or even Newton-Raphson 
method) should be applied in order to obtain approximate numerical values for $\mathcal{S}_{\delta U}^{(0)}$ (this statement is supported by the Abel-Galois irreducibility 
theorem \cite{Galois}, which asserts that polynomial equations of degree $\geq 5$ do not produce, in general, algebraic solutions, only numerical solutions).
\begin{table}[t]
\caption{Numerical values for $\{ \mathfrak{R}_{i} \}_{1 \leq i \leq 4}$ considering the ground state $\{ | 0 \rgg_{N} \}_{2 \leq N \leq 6}$ described in Appendix 
A. It is important to stress that such results were inferred from their respective exact algebraic counterparts, which illustrate, in principle, not only the 
hierarchical relation depicted in Eq. (\ref{s3e21}) but also the quantum-algebraic framework developed in the previous sections. In addition, note that eigenvalues
extracted from the Harper Hamiltonian for $N > 6$ do not yield easy-to-compute expressions for the tightest bounds $\mathfrak{R}_{1}$, this fact being supported
by Abel-Galois irreducibility theorem for polynomial equations.}
\centering
\begin{tabular}{llllll}
\hline
$N$             & $\mathfrak{R}_{1}$ & $\mathfrak{R}_{2}$ & $\mathfrak{R}_{3}$ & $\mathfrak{R}_{4}$ \\ \hline
$2$             & $1$                & $1$                & $1$                & $1$                \\
$3$             & $\approx 3.254$    & $\approx 2.182$    & $\approx 1.895$    & $\approx 1.866$    \\
$4$             & $4$                & $3$                & $2$                & $2$                \\
$5$             & $\approx 3.781$    & $\approx 3.469$    & $\approx 2.987$    & $\approx 1.951$    \\ 
$6$             & $\approx 3.348$    & $\approx 1.915$    & $\approx 1.915$    & $\approx 1.866$    \\ \hline
\end{tabular}
\end{table}

\section{The connection with discrete Weyl function}

How the discrete Weyl function \cite{Ferrie} can be employed to measure a particular family of expectation values --- for instance, $\lgg {\bf U}^{\alf} 
{\bf V}^{\bet} \rgg$ for $\{ \alf,\bet \} \in \mathbb{Z}_{N}$ --- here mapped into finite-dimensional discrete phase spaces? To answer this specific 
question, we initially recall certain basic mathematical tools which correspond to the central core of that theoretical formulation presented in Ref. \cite{MR1}. 
This procedure will lead us to establish a parallel quantum-algebraic framework for those results obtained in Section 3, whose connection with the discrete 
Weyl function represents a first step towards effective experimental measurements via tomographic reconstructions of finite quantum states \cite{Ap1}. 
Throughout this section, we will assume $N$ odd.\footnote{For completeness reasons, it is important to stress that even dimensionalities can also be 
dealt with simply by working on non-symmetrized intervals.}

The particular set of $N^{2}$ operators $\{ \bDelta(\mu,\nu) \}_{\mu,\nu=-\ell,\ldots,\ell}$ characterizes a complete orthonormal unitary operator basis 
which leads us to construct all possible kinematical and/or dynamical quantities belonging to a given $N$-dimensional state vector space. For instance, the 
decomposition of any linear operator $\mathbf{O}$ in this basis assumes the expression
\be
\lb{s5e22}
\mathbf{O} = \frac{1}{N} \sum_{\mu,\nu = - \ell}^{\ell} \mathscr{O}(\mu,\nu) \bDelta(\mu,\nu) ,
\ee
where $\mathscr{O}(\mu,\nu) \equiv \mathrm{Tr} [ \bDelta(\mu,\nu) \mathbf{O} ]$ represent coefficients evaluated through trace operation and
\be
\lb{s5e23} 
\bDelta(\mu,\nu) \coloneq \frac{1}{N} \sum_{\eta,\xi = - \ell}^{\ell} \om^{-(\eta \nu - \xi \mu)} \mathbf{D}(\eta,\xi)
\ee
defines the aforementioned operator basis here expressed in terms of a discrete Fourier transform of the displacement operator
\bd
\mathbf{D}(\eta,\xi) = \om^{- \{ 2^{-1} \eta \xi \} } \sum_{\gam = - \ell}^{\ell} \om^{\gam \eta} | u_{\gam} \rgg \lgg u_{\gam - \xi} | =
\om^{- \{ 2^{-1} \eta \xi \} } \sum_{\gam = - \ell}^{\ell} \om^{- \gam \xi} | v_{\gam + \eta} \rgg \lgg v_{\gam} | .
\ed
In such a case, note that $\om^{- \{ 2^{-1} \eta \xi \} }$ consists of a specific phase whose argument satisfies the mathematical rule
$2 \{ 2^{-1} \eta \xi \} = \eta \xi + kN$ for all $k \in \mathbb{Z}_{N}$; furthermore, note that these labels assume integer values in the symmetric 
interval $[-\ell,\ell]$ with $\ell = \frac{N-1}{2}$ fixed.

According to expansion (\ref{s5e22}), the decomposition of any density operator $\ro$ in the mod($N$)-invariant unitary operator basis (\ref{s5e23}) has as
coefficients the discrete Wigner function $\mathscr{W}_{\rho}(\mu,\nu) \coloneq \mathrm{Tr} [ \bDelta(\mu,\nu) \ro ]$, which leads us, in principle, to establish 
an analytical expression for the mean value $\lgg \mathbf{O} \rgg \equiv \mathrm{Tr} [ \mathbf{O} \ro ]$, that is
\be
\lb{s5e24}
\lgg \mathbf{O} \rgg = \frac{1}{N} \sum_{\mu,\nu = - \ell}^{\ell} \mathscr{O}(\mu,\nu) \mathscr{W}_{\rho}(\mu,\nu) .
\ee
In what concerns to $\mathscr{W}_{\rho}(\mu,\nu)$, it is particularly worth mentioning that such a function is connected to the Weyl function
$\widetilde{\mathscr{W}}_{\rho}(\eta,\xi) \coloneq \mathrm{Tr} [ \mathbf{D}(\eta,\xi) \ro ]$ by means of a mere discrete Fourier transform, and its complexity
basically depends on the initial quantum state adopted for the physical system under investigation. 

To illustrate the mathematical steps used in the evaluation of $\mathscr{O}(\mu,\nu)$, let us consider those specific combinations of cosine and sine operators 
exhibited in Section 3, as well as the intermediate result
\bd
\mathbf{D}^{\dagger}(\eta,\xi) \mathbf{U}^{\alf} \mathbf{V}^{\bet} \mathbf{D}(\eta,\xi) = \om^{\alf \xi + \bet \eta} \mathbf{U}^{\alf} \mathbf{V}^{\bet}
\ed
since the trace operation 
\bd
\mathrm{Tr} [ \mathbf{D}(\eta,\xi) \mathbf{U}^{\alf} \mathbf{V}^{\bet} ] = N \om^{- \{ 2^{-1} \eta \xi \} + \eta \bet} \delta_{\eta + \alf,0}^{[ N ]} \,
\delta_{\xi - \bet,0}^{[ N ]} 
\ed
will be necessary in the next steps. Thus, after some repetitive calculations, we achieve the set of formal expressions
\brr
\!\!\!\!\! \mathrm{Tr} [ \mathbf{D}(\eta,\xi) \mathbf{C}_{\innu} \mathbf{C}_{\innv} ] &=& \frac{N}{4} \, \om^{- \{ 2^{-1} \eta \xi \} } \lpar 
\delta_{\eta + 1, 0}^{[ N ]} + \delta_{\eta - 1, 0}^{[ N ]} \rpar \lpar \om^{- \eta} \delta_{\xi + 1, 0}^{[ N ]} + \om^{\eta} 
\delta_{\xi - 1, 0}^{[ N ]} \rpar \nn \\
\!\!\!\!\! \mathrm{Tr} [ \mathbf{D}(\eta,\xi) \mathbf{C}_{\innu} \mathbf{S}_{\innv} ] &=& \mathrm{i} \, \frac{N}{4} \, \om^{- \{ 2^{-1} \eta \xi \} } \lpar 
\delta_{\eta + 1, 0}^{[ N ]} + \delta_{\eta - 1, 0}^{[ N ]} \rpar \lpar \om^{- \eta} \delta_{\xi + 1, 0}^{[ N ]} - \om^{\eta} 
\delta_{\xi - 1, 0}^{[ N ]} \rpar \nn \\
\!\!\!\!\! \mathrm{Tr} [ \mathbf{D}(\eta,\xi) \mathbf{S}_{\innu} \mathbf{C}_{\innv} ] &=& - \mathrm{i} \, \frac{N}{4} \, \om^{- \{ 2^{-1} \eta \xi \} } \lpar 
\delta_{\eta + 1, 0}^{[ N ]} - \delta_{\eta - 1, 0}^{[ N ]} \rpar \lpar \om^{- \eta} \delta_{\xi + 1, 0}^{[ N ]} + \om^{\eta} 
\delta_{\xi - 1, 0}^{[ N ]} \rpar \nn \\
\!\!\!\!\! \mathrm{Tr} [ \mathbf{D}(\eta,\xi) \mathbf{S}_{\innu} \mathbf{S}_{\innv} ] &=& \frac{N}{4} \, \om^{- \{ 2^{-1} \eta \xi \} } \lpar 
\delta_{\eta + 1, 0}^{[ N ]} - \delta_{\eta - 1, 0}^{[ N ]} \rpar \lpar \om^{- \eta} \delta_{\xi + 1, 0}^{[ N ]} - \om^{\eta} 
\delta_{\xi - 1, 0}^{[ N ]} \rpar \nn
\err
which represents, in such a case, the dual counterpart of $\mathscr{O}(\mu,\nu)$ --- the superscript $[N]$ on the Kronecker delta denotes that this
function is different from zero when its discrete labels are congruent modulo $N$. 

The second and last step consists in computing, for each case above, its respective discrete Fourier transform, which basically depends on the
intermediate result
\bd
\mathrm{Tr} [ \bDelta(\mu,\nu) \mathbf{U}^{\alf} \mathbf{V}^{\bet} ] = \om^{\bet \mu + \alf \nu - \alf \bet + \{ 2^{-1} \alf \bet \}} .
\ed
So, the final expressions admit the simple forms
\brr
\!\!\!\!\!\!\!\!\!\! \lpar \mathbf{C}_{\innu} \mathbf{C}_{\innv} \rpar (\mu,\nu) &=& \frac{1}{4} \lbk \lpar \om^{\mu - \nu} + \om^{- \mu + \nu} \rpar 
\om^{1 - \{ 2^{-1} \} } + \lpar \om^{\mu + \nu} + \om^{- \mu - \nu} \rpar \om^{- 1 + \{ 2^{-1} \} } \rbk \nn \\
\!\!\!\!\!\!\!\!\!\! \lpar \mathbf{C}_{\innu} \mathbf{S}_{\innv} \rpar (\mu,\nu) &=& - \frac{\mathrm{i}}{4} \lbk \lpar \om^{\mu - \nu} - \om^{- \mu + \nu} \rpar 
\om^{1 - \{ 2^{-1} \} } + \lpar \om^{\mu + \nu} - \om^{- \mu - \nu} \rpar \om^{- 1 + \{ 2^{-1} \} } \rbk \nn \\
\!\!\!\!\!\!\!\!\!\! \lpar \mathbf{S}_{\innu} \mathbf{C}_{\innv} \rpar (\mu,\nu) &=& \frac{\mathrm{i}}{4} \lbk \lpar \om^{\mu - \nu} - \om^{- \mu + \nu} \rpar 
\om^{1 - \{ 2^{-1} \} } - \lpar \om^{\mu + \nu} - \om^{- \mu - \nu} \rpar \om^{- 1 + \{ 2^{-1} \} } \rbk \nn \\
\!\!\!\!\!\!\!\!\!\! \lpar \mathbf{S}_{\innu} \mathbf{S}_{\innv} \rpar (\mu,\nu) &=& \frac{1}{4} \lbk \lpar \om^{\mu - \nu} + \om^{- \mu + \nu} \rpar 
\om^{1 - \{ 2^{-1} \} } - \lpar \om^{\mu + \nu} + \om^{- \mu - \nu} \rpar \om^{- 1 + \{ 2^{-1} \} } \rbk \nn
\err
where $\lpar {\bf A} {\bf B} \rpar (\mu,\nu) \equiv \mathrm{Tr} \lbk \bDelta (\mu,\nu) {\bf A} {\bf B} \rbk$ denotes the above mapped expressions. Consequently, with 
the aid of Eq. (\ref{s5e24}), the associated mean values can be promptly obtained: for instance, if one considers $\lgg \mathbf{U}^{\alf} \mathbf{V}^{\bet} \rgg$, 
we easily reach the important result
\be
\lb{s5e25} 
\lgg \mathbf{U}^{\alf} \mathbf{V}^{\bet} \rgg = \om^{- \alf \bet + \{ 2^{-1} \alf \bet \}} \widetilde{\mathscr{W}}_{\rho}(\alf,-\bet) .
\ee
Now, let us pay attention to $\lgg \mathbf{C}_{\innu} \mathbf{C}_{\innv} \rgg$ and its particular link with the discrete Weyl function,
\brr
\lb{s5e26}
\lgg \mathbf{C}_{\innu} \mathbf{C}_{\innv} \rgg &=& \frac{1}{4} \lbk \widetilde{\mathscr{W}}_{\rho}(-1,-1) + \widetilde{\mathscr{W}}_{\rho}(1,1) \rbk
\om^{1 - \{ 2^{-1} \} } \nn \\
& & + \frac{1}{4} \lbk \widetilde{\mathscr{W}}_{\rho}(1,-1) + \widetilde{\mathscr{W}}_{\rho}(-1,1) \rbk \om^{-1 + \{ 2^{-1} \} } .
\err
This remarkable result states that, for a given $\ro$ and its respective discrete Weyl function $\widetilde{\mathscr{W}}_{\rho}(\eta,\xi)$, $\lgg \mathbf{C}_{\innu}
\mathbf{C}_{\innv} \rgg$ can be promptly inferred from tomography measures upon such a function at specific points of the dual $N$-dimensional discrete phase space.
Besides, if one applies the triangle inequality for $| \lgg \mathbf{C}_{\innu} \mathbf{C}_{\innv} \rgg |$, both the upper and lower bounds can be easily established 
in this context,\footnote{Let $\mathbf{A}$ and $\mathbf{B}$ characterize two general matrices of same size, as well as $\ro$ denote the density matrix related to a
physical system described by a finite-dimensional state vector space. The additional inequality $| \mathrm{Tr} [ \mathbf{A} \mathbf{B} \ro ] |^{2} \leq \mathrm{Tr} 
[ \mathbf{A} \mathbf{A}^{\dagger} \ro ] \mathrm{Tr} [ \mathbf{B} \mathbf{B}^{\dagger} \ro ]$ (see Ref. \cite[page 230]{Ingemar}) establishes a new upper bound for 
Eq. (\ref{s5e26}) since $| \lgg \mathbf{C}_{\innu} \mathbf{C}_{\innv} \rgg |^{2} \leq \lgg \mathbf{C}_{\innu}^{2} \rgg \lgg \mathbf{C}_{\innv}^{2} \rgg$, where
\bd
\lgg \mathbf{C}_{\innu}^{2} \rgg = \frac{1}{2} + \frac{1}{4} \lbk \widetilde{\mathscr{W}}_{\rho}(2,0) + \widetilde{\mathscr{W}}_{\rho}(-2,0) \rbk 
\ed
and 
\bd
\lgg \mathbf{C}_{\innv}^{2} \rgg = \frac{1}{2} + \frac{1}{4} \lbk \widetilde{\mathscr{W}}_{\rho}(0,2) + \widetilde{\mathscr{W}}_{\rho}(0,-2) \rbk .
\ed
This complementary result basically represents a step forward in our comprehension on hierarchical relations involving those means values listed in Remark 4.} 
\bd
| \lgg \mathbf{C}_{\innu} \mathbf{C}_{\innv} \rgg | \lesseqqgtr \frac{1}{4} \lpar | \widetilde{\mathscr{W}}_{\rho}(-1,-1) + 
\widetilde{\mathscr{W}}_{\rho}(1,1) | \pm | \widetilde{\mathscr{W}}_{\rho}(1,-1) + \widetilde{\mathscr{W}}_{\rho}(-1,1) | \rpar .
\ed
Similar procedure leads us to prove that $| \lgg \mathbf{S}_{\innu} \mathbf{S}_{\innv} \rgg |$ shares exactly the same bounds, while the remaining quantities
are bounded by the inequalities
\brr
\lbr | \lgg \mathbf{C}_{\innu} \mathbf{S}_{\innv} \rgg | , | \lgg \mathbf{S}_{\innu} \mathbf{C}_{\innv} \rgg | \rbr & \lesseqqgtr & \frac{1}{4} \lpar 
| \widetilde{\mathscr{W}}_{\rho}(-1,-1) - \widetilde{\mathscr{W}}_{\rho}(1,1) | \rpar \nn \\
& & \pm \frac{1}{4} \lpar | \widetilde{\mathscr{W}}_{\rho}(1,-1) - \widetilde{\mathscr{W}}_{\rho}(-1,1) | \rpar . \nn
\err
As a last note, let us mention that $\mathscr{V}_{\innu}$ and $\mathscr{V}_{\innv}$ can also be expressed in terms of specific Weyl functions with the help 
of Eq. (\ref{s5e25}), \ie, $\mathscr{V}_{\innu} = 1 - | \widetilde{\mathscr{W}}_{\rho}(1,0) |^{2}$ and $\mathscr{V}_{\innv} = 1 - | \widetilde{\mathscr{W}}_{\rho}
(0,-1) |^{2}$.

In particular, it is worth stressing that the compilation of results obtained in this section allows us to rewrite inequality (\ref{s3e13}) into a new quantum-algebraic 
framework, since $\mathcal{F}(\mathbf{U},\mathbf{V})$ and $\mathcal{H}(\mathbf{U},\mathbf{V})$ now depend on specific combinations of discrete Weyl functions. From an
experimental point of view, this observation represents an effective gain towards tomographic measurements in $N$-dimensional discrete phase spaces of such a generalized
inequality for any finite quantum state $\ro$ \cite{Ap1}.

\section{Concluding remarks}

Through a well succeeded concatenation of efforts in a recent past \cite{Ferrie,MR1}, we have made great advances (from a theoretical point of view) in 
constructing certain sound quantum-algebraic frameworks for finite-dimensional discrete phase spaces. Since unitary operators represent the basic constituent 
blocks of these theoretical frameworks (in particular, the Schwinger's approach for unitary operators \cite{Schwinger}), let us focus our attention on some 
real and effective gains obtained from this paper whose formal implications deserve to be carefully discussed.  

\begin{itemize}
\item The Massar-Spindel inequality does not explain the tightest bounds exhibited in Fig. 1 for the product $\mathcal{S}_{\delta \innu} \mathcal{S}_{\delta \innv}$
when dimensions $N \geq 3$ are considered in the numerical evaluations. This concrete evidence suggests the implementation of an effective search for different
inequalities and new finite quantum states whose implications lead us not only to obtain a reasonable set of improved mathematical results, but also to answer
certain important questions that emerge from such an evidence. 
\item The mathematical background for achieving different inequalities related to the aforementioned unitary operators has the RS uncertainty principle 
as solid starting-point \cite{MR1}. The initial algebraic advantage of this investigative approach is the inclusion of that contribution coming from the
anticommutation relation between two Hermitian non-commuting operators connected via discrete Fourier transform (and/or also related through the Pontryagin 
duality \cite{Rudin}). In fact, this particular contribution allows to include, into the algebraic approach, some additional terms --- here associated with 
the mean values of different products of the cosine and sine operators --- which are responsible for correlations between the unitary operators. In principle,
the theorem derived from this constructive process represents a first important point to be strongly emphasized since Eq. (\ref{s3e13}) introduces a tighter 
bound whose mathematical properties depend on the $N$-dimensional state vector space in which the initial quantum state is defined.
\item The hierarchical relation (\ref{s3e21}) derived from the tighter bound characterizes a solid bridge between two `distant' bounds: the first one
consists of a zeroth-order approximation that confirms the Massar-Spindel result \cite{Massar} (however, it does not depend on the initial quantum state
or even on the dimension which the state vector space is embedded, these facts being considered as a severe limitation for their result); while the second 
one describes the tightest bound for the left-hand side of Eq. (\ref{s3e13}) and depicts the importance of the initial quantum state for a given dimension
$N$. This simple (but important) observation justifies our search for finite ground states which quantitatively describe those numerical values obtained in
Fig. 1 for the tightest bounds. 
\item How to construct a finite quantum state which formally explains the tightest bounds verified in the numerical calculations? To answer this fundamental
question, it is necessary to establish an adequate mathematical prescription that leads us to obtain a set of normalized eigenfunctions which constitutes a
complete orthonormal basis in a $N$-dimensional state vector space. In this sense, Appendix A deals with such a task presenting a reliable mathematical
relation between Harper functions and tightest bounds by means of ground states $\{ | 0 \rgg_{N} \}$ specifically constructed for dimensions $N \in [2,6]$.
These finite quantum states indeed describe perfectly all the numerical values exhibited in Fig. 1 for the tightest bounds, and illustrate the hierarchical 
relation (\ref{s3e21}) as well --- see Table 1. 
\item The bounds determined in this paper for the product $\mathcal{S}_{\delta \innu} \mathcal{S}_{\delta \innv}$ exhibit a special link with the theoretical
formulation of finite-dimensional discrete phase spaces through the discrete Weyl funtion. In fact, this connection establishes an interesting link between 
both the Schwinger and Weyl prescriptions for unitary operators, which leads us to guess on the possibility of experimental observation with the help of 
tomographic measurements.
\end{itemize}

Now, let us discuss some pertinent points associated with the uncertainty principle for Schwinger unitary operators. The first point concerns the 
Harper functions and their remarkable connection with the tightest bounds through the ground states $\{ | 0 \rgg_{N} \}$ for a given dimension $N$ fixed.
Despite the worked-examples in Appendix A belonging to the closed interval $2 \leq N \leq 6$, this fact does not represent any apparent limitation related
to the quantum-algebraic framework here exposed. In fact, these results consist of a solid starting point for a future search of $\{ \lgg u_{\alf} | n 
\rgg \}_{0 \leq n \leq N-1}$ with $\alf \in \mathbb{Z}_{N}$, whose general expression will correspond to a new paradigm for finite quantum states with
immediate implications in the Fourier analysis on finite groups \cite{Terras,Rudin} (and/or finite fields \cite{Lidl}), as well as in the analysis of
signal processing \cite{Dickinson,Astola}. This particular task is currently in progress and the results will be presented in elsewhere.

The second point focus on the systematic study recently developed in Ref. \cite{Wang} for property testing of unitary operators, where $D^{2}({\bf U},{\bf V}) 
\coloneq 1 - \frac{1}{N} | \mathrm{Tr} [ {\bf U}^{\dagger} {\bf V} ] |$ represents a `normalized distance measure that reflects the average difference between 
unitary operators'. With respect to this specific measure, Wang shows that both the Clifford and orthogonal groups can be efficiently tested through algorithms 
with intrinsic mathematical virtues (namely, query complexities independent of the system's size and one-sided error). Since the results here obtained describe 
an uncertainty principle for unitary operators, it seems reasonable to investigate how these different --- but complementary --- approaches can be juxtaposed in 
order to produce a unified framework for determined tasks in quantum information theory \cite{Books}. 

As a final comment, let us briefly mention that our results also touch on some fundamental questions inherent to quantum mechanics (such as spin-squeezing
and entanglement effects \cite{MGD}), discrete fractional Fourier transform \cite{Cotfas}, and generalized uncertainty principle into the quantum-gravity 
context \cite{MR1}.

\ack 

The authors thank Di\'{o}genes Galetti for helpful discussions and comments on highly pertinent questions related to this work. 

\appendix
\section{Harper's equation, quantum Fourier transform, tightest bound and their inherent connections with unitary operators}

\begin{pf*}{{\bf Definition (Harper functions).}}
Let $\lbr | n \rgg \rbr_{0 \leq n \leq N-1}$ describe a particular set of eigenvectors defined in a $N$-dimensional state vector space that 
simultaneously diagonalizes both the Hamiltonian $({\bf H})$ and Fourier $(\fou)$ operators, that is, ${\bf H} | n \rgg = h_{n} | n \rgg$ and 
$\fou | n \rgg = f_{n} | n \rgg$ for a given dimension $N$ fixed. In such a case, $\lbr h_{n},f_{n} \rbr_{0 \leq n \leq N-1}$ represents the 
corresponding set of eigenvalues related to the respective Hamiltonian and Fourier operators. Since 
\bd
{\bf H} \fou | n \rgg = \fou {\bf H} | n \rgg = f_{n} h_{n} | n \rgg \Rightarrow \lbk {\bf H},\fou \rbk | n \rgg = 0 | n \rgg ,
\ed
the intrinsic mathematical properties associated with the discrete representation $\lbr | u_{\alf} \rgg \rbr_{0 \leq \alf \leq N-1}$ allow to obtain 
the general equation 
\be
\lb{a1}
\lgg u_{\alf} | {\bf H} \fou | n \rgg = f_{n} h_{n} \lgg u_{\alf} | n \rgg , 
\ee
whose solution set $\lbr \lgg u_{\gam} | n \rgg \rbr \in \mathbb{R}$ yields a complete orthonormal basis of real eigenfunctions genuinely labelled 
by discrete variables. The Harper functions are here attained when one considers ${\bf H}$ as being a Harper-type operator \cite{Dickinson}.
\end{pf*}

As a first application, let us, for now, adopt the discrete Fourier operator\footnote{For $N$ odd and discrete labels assuming integer values in the
symmetric interval $[ - \ell,\ell ]$ with $\ell = \frac{N-1}{2}$ fixed, it is worth stressing that $\fou^{2}$ coincides with that parity operator
$\opp$ previously defined in Ref. \cite{MR1}.}
\bd
\fou \coloneq \sum_{\bet = 0}^{N-1} | v_{\bet} \rgg \lgg u_{\bet} | = \frac{1}{\sqrt{N}} \sum_{\bet,\bet^{\prime} = 0}^{N-1} \om^{\bet \bet^{\prime}}
| u_{\bet^{\prime}} \rgg \lgg u_{\bet} | \; \Rightarrow \; \fou \fou^{\dagger} = \fou^{\dagger} \fou = {\bf 1},
\ed
as well as that Harper Hamiltonian operator previously discussed in Section 4, namely, ${\bf H} = - \sin (\th) {\bf C}_{\innu} - \cos (\th) {\bf C}_{\innv}$ 
for $\th \in \lbk 0, \frac{\pi}{2} \rbk$. Thus, Eq. (\ref{a1}) assumes the functional form
\be
\lb{a2}
\sum_{\bet = 0}^{N-1} \mathds{O}(\alf,\bet;N) \lgg u_{\bet} | n \rgg = f_{n} h_{n} \lgg u_{\alf} | n \rgg ,
\ee
where 
\bd
\mathds{O}(\alf,\bet;N) \equiv \lgg u_{\alf} | {\bf H} | v_{\bet} \rgg = - \lbk \sin (\th) \cos \lpar \frac{2 \pi \alf}{N} \rpar + \cos (\th) \cos \lpar
\frac{2 \pi \bet}{N} \rpar \rbk \lgg u_{\alf} | v_{\bet} \rgg
\ed
represents the mapped expression of the Hamiltonian operator in the discrete representations $\lbr | u_{\alf} \rgg, | v_{\bet} \rgg \rbr$ with 
$\lgg u_{\alf} | v_{\bet} \rgg = \frac{1}{\sqrt{N}} \om^{\alf \bet}$ and $\om = \exp \lpar \frac{2 \pi \mathrm{i}}{N} \rpar$; besides, 
$\lbr \lgg u_{\gam} | n \rgg \rbr \in \mathbb{R}$ denotes the Harper functions for a given $N \in \mathbb{N}^{\ast}$. In fact, such a result can also be 
split up into two combined equations as follows: 
\be
\lb{a3}
\sin (\th) \cos \lpar \frac{2 \pi \alf}{N} \rpar \lgg u_{\alf} | n \rgg + \frac{1}{2} \cos (\th) \lpar \lgg u_{\alf - 1} | n \rgg +
\lgg u_{\alf + 1} | n \rgg \rpar = - h_{n} \lgg u_{\alf} | n \rgg
\ee
and
\be
\lb{a4}
\frac{1}{\sqrt{N}} \sum_{\bet = 0}^{N-1} \om^{\alf \bet} \lgg u_{\bet} | n \rgg = f_{n} \lgg u_{\alf} | n \rgg .
\ee
The first one describes a three-term recurrence relation and also depicts the well-known Harper's equation, whose link with discrete harmonic oscillator 
and discrete fractional Fourier transform was already discussed by Barker and coworkers \cite{Barker}; whilst the second one reflects exactly the eigenvalue 
problem investigated by Mehta \cite{Mehta} when $f_{n} = \mathrm{i}^{n}$ (in this particular case, see Ref. \cite{Terras} for supplementary material), although 
his ansatz solution
\bd
\lgg u_{\alf} | n \rgg = \mathcal{N}_{n} \frac{(-\mathrm{i})^{n}}{\sqrt{N}} \sum_{\kappa = - \infty}^{\infty} \exp \lpar - \frac{\pi}{N} \kappa^{2} +
\frac{2 \pi \mathrm{i}}{N} \kappa \alf \rpar \mathit{H}_{n} \lpar \sqrt{\frac{2 \pi}{N}} \kappa \rpar
\ed
does not represent a complete set of orthonormal eigenfunctions \cite{Ruzzi} --- in such ansatz solution, $\mathcal{N}_{n}$ corresponds to a normalization constant
and $\mathit{H}_{n}(z)$ denotes the Hermite polynomials.

Summarizing, Eq. (\ref{a3}) yields, in general, a set of real eigenvalues $\{ h_{n} \}$ whose respective eigenfunctions $\{ \lgg u_{\alf} | n \rgg \}$ 
constitute a complete orthonormal basis in a $N$-dimensional state vector space. In this specific case, both the eigenvalues and eigenfunctions are dependent 
on the angle variable $\th \in \lbk 0,\frac{\pi}{2} \rbk$, which leads us to determine the maximum of $\cos (\th) | \lgg {\bf U} \rgg | + \sin (\th) | \lgg 
{\bf V} \rgg |$\footnote{Such a quantity implicitly defines the boundary --- or, more precisely, the convex hull --- of the accessible region related to the
$\{ | \lgg {\bf U} \rgg |, | \lgg {\bf V} \rgg | \}$-space \cite{Massar}.} for each particular situation $h_{n} \rightleftharpoons \lgg u_{\alf} | n \rgg$ 
with $N$ fixed. Thus, the global maximum obtained from this mathematical procedure allows not only to fix a given value of $\th$, but also to estimate the 
smallest eigenvalue of the Hermitian operator ${\bf H}$; consequently, the corresponding eigenvector will describe the ground state $| 0 \rgg$ characterized 
by $\th = \frac{\pi}{4}$ and $f_{0} = +1$ for any dimension $N$ (it is worth stressing that theoretical and numerical calculations confirm these results).
Next, let us consider the $N = 2,\ldots,6$ cases for $\th = \frac{\pi}{4}$ fixed, in order to provide a complete list of results exhibited in Table 1 associated
with the normalized ground state.
\begin{itemize}
\item Case $N=2$ (prime dimension) and $h_{0} = -1$. This first example obeys the criterion ``easy to calculate", once the corresponding ground state
\bd
| 0 \rgg_{2} = \frac{\sqrt{2 + \sqrt{2}}}{2} | u_{0} \rgg + \frac{\sqrt{2 - \sqrt{2}}}{2} | u_{1} \rgg
\ed
allows to obtain $\mathscr{V}_{\innu (\innv)} = \frac{1}{2}$ and $\mathscr{V}_{\delta \innu (\delta \innv)} = 1$, which implies that
$\mathcal{S}^{(0)}_{\delta \innu} = 1$.

\item Case $N=3$ (prime dimension) and $h_{0} = - \frac{\sqrt{6} + \sqrt{2}}{4}$. This particular example yields certain interesting peculiarities
since its respective ground state
\bd
| 0 \rgg_{3} = \sqrt{\frac{3 + \sqrt{3}}{6}} | u_{0} \rgg + \sqrt{\frac{3 - \sqrt{3}}{12}} | u_{1} \rgg + \sqrt{\frac{3 - \sqrt{3}}{12}} | u_{2} \rgg
\ed
leads to achieve $\mathscr{V}_{\innu (\innv)} = \frac{1}{2} + \frac{2 - \sqrt{3}}{8}$ and $\mathscr{V}_{\delta \innu (\delta \innv)} = 15 - 8 \sqrt{3}$, 
which corroborate those results obtained by Opatrn\'{y} \cite{Jackiw} for the number and phase operators. Furthermore, it is easy to demonstrate that
$\mathcal{S}^{(0)}_{\delta \innu} = 7 - 3 \sqrt{3}$, which justifies the approximated numerical value $\mathfrak{R}_{1} \approx 3.254$ exhibited in Table 1.

\item Case $N=4$ (even dimension) and $h_{0}=-1$. In such a situation, the ground state
\bd
| 0 \rgg_{4} = \frac{2 + \sqrt{2}}{4} | u_{0} \rgg + \frac{\sqrt{2}}{4} | u_{1} \rgg + \frac{2 - \sqrt{2}}{4} | u_{2} \rgg + \frac{\sqrt{2}}{4} | u_{3} \rgg
\ed
attains those same values of $\mathscr{V}_{\innu (\innv)}$ and $\mathscr{V}_{\delta \innu (\delta \innv)}$ verified in $N=2$; however, it is worth stressing
that $\mathcal{S}^{(0)}_{\delta \innu} = 2$, since $\eps = \frac{1}{2}$ in this particular case.

\item Case $N=5$ (prime dimension) and $h_{0} = - \frac{\sqrt{2} + \sqrt{10} + 2 \sqrt{35 + \sqrt{5}}}{16}$. For this nontrivial example, the normalized ground 
state assumes the form
\bd
| 0 \rgg_{5} = \frac{1}{\sqrt{1 + 2( \mathfrak{a}_{1}^{2} + \mathfrak{a}_{2}^{2} )}} \lpar | u_{0} \rgg + \mathfrak{a}_{1} | u_{1} \rgg + 
\mathfrak{a}_{2} | u_{2} \rgg + \mathfrak{a}_{2} | u_{3} \rgg + \mathfrak{a}_{1} | u_{4} \rgg \rpar
\ed
with $\mathfrak{a}_{1} = \frac{\sqrt{5} + \sqrt{2(35 + \sqrt{5})} - 7}{8}$ and $\mathfrak{a}_{2} = \frac{3 \sqrt{5} - \sqrt{2(35 + \sqrt{5})} + 3}{8}$. So,
after some lengthy calculations, the exact expression $\mathscr{V}_{\delta \innu (\delta \innv)} = \frac{45 - \sqrt{5} - \sqrt{110 + 38 \sqrt{5}}}{19 + 
\sqrt{5} + \sqrt{110 + 38 \sqrt{5}}}$ for the variance allows to show that
\bd
\mathcal{S}^{(0)}_{\delta \innu} = \frac{\lpar 5 + \sqrt{25 + 10 \sqrt{5}} \, \rpar \lpar 45 - \sqrt{5} - \sqrt{110 + 38 \sqrt{5}} \, \rpar}
{5 \lpar 19 + \sqrt{5} + \sqrt{110 + 38 \sqrt{5}} \, \rpar} \approx 1.9444 ,
\ed
corroborating, in this way, the numerical result $\mathfrak{R}_{1} \approx 3.781$ (see Table 1).

\item Case $N=6$ (even dimension) and $h_{0} = - \frac{\sqrt{10 + 2 \sqrt{21}}}{4}$. Similarly to the previous case, in this situation the ground state admits
\bd
| 0 \rgg_{6} = \frac{1}{\sqrt{1 + 2( \mathfrak{b}_{1}^{2} + \mathfrak{b}_{2}^{2} ) + \mathfrak{b}_{3}^{2}}} \lpar | u_{0} \rgg + \mathfrak{b}_{1} | u_{1} \rgg + 
\mathfrak{b}_{2} | u_{2} \rgg + \mathfrak{b}_{3} | u_{3} \rgg + \mathfrak{b}_{2} | u_{4} \rgg + \mathfrak{b}_{1} | u_{5} \rgg \rpar
\ed
as a solution of Eq. (\ref{a2}), where
\brr
\mathfrak{b}_{1} &=& \frac{-2 + \sqrt{5 + \sqrt{21}}}{2} = \frac{-4 + \sqrt{6} + \sqrt{14}}{4} \nn \\ 
\mathfrak{b}_{2} &=& \frac{5 + \sqrt{21} - 3 \sqrt{5 + \sqrt{21}}}{2} = \frac{10 - 3 \sqrt{6} - 3 \sqrt{14} + 2 \sqrt{21}}{4} \nn \\ 
\mathfrak{b}_{3} &=& -4 + \sqrt{6} - \sqrt{21} + 2 \sqrt{5 + \sqrt{21}} = -4 + 2 \sqrt{6} + \sqrt{14} - \sqrt{21} \nn
\err
represent the respective coefficients. Note that $\mathscr{V}_{\delta \innu (\delta \innv)} = 19 - 4 \sqrt{21}$, which implies in 
$\mathcal{S}^{(0)}_{\delta \innu} = 19 ( 1 + \sqrt{3} ) - 4 \sqrt{42 ( 2 + \sqrt{3} )} \approx 1.8297$ (this result justifies that numerical value
$\mathfrak{R}_{1} \approx 3.348$ appeared in Table 1 for $N=6$).
\end{itemize}

As an initial purpose, these first theoretical results related to the ground state are sufficient to clarify the numerical results depicted in Fig. 1
for the tightest bound. In fact, the results exhibited in this appendix indeed represent a first investigative step towards a general mathematical
recipe that presents as a primary product the eigenfunctions $\{ \lgg u_{\alf} | n \rgg \}_{0 \leq n \leq N-1}$, which differ from that Mehta's ansatz
solution.


\end{document}